\documentclass[lettersize,journal]{IEEEtran}
\usepackage{amsmath,amssymb,bm}
\usepackage{array}
\usepackage{booktabs}
\usepackage{multirow}
\usepackage{cite}
\usepackage{url}
\usepackage{enumitem}

\newcommand{\Vset}{\mathcal{V}}
\newcommand{\Eset}{\mathcal{E}}

\newcommand{\Tset}{\mathcal{T}}

\usepackage{graphicx}
\usepackage{amsmath,amssymb}
\usepackage{booktabs}
\usepackage{array}
\usepackage[linesnumbered,ruled,vlined]{algorithm2e}
\usepackage{mathtools}

\usepackage[textsize=scriptsize]{todonotes} 
\begin{document}
	
	\title{Lightweight Onboard Routing and Topology Co-Construction for Autonomous GNSSs}
	
	\author{
		\IEEEauthorblockN{
			Huan Yan\IEEEauthorrefmark{1},
			Juan A. Fraire\IEEEauthorrefmark{2}\IEEEauthorrefmark{3},
			Rongqin Yang\IEEEauthorrefmark{1},
			Wenfeng Li\IEEEauthorrefmark{1},
			and Kanglian Zhao\IEEEauthorrefmark{1}
		}
		
		\IEEEauthorblockA{
			\IEEEauthorrefmark{1} Nanjing University, Nanjing, China \\
			\IEEEauthorrefmark{2} Inria, INSA Lyon, CITI, UR3720, 69621 Villeurbanne, France \\
			\IEEEauthorrefmark{3} CONICET -- Universidad Nacional de Córdoba, Córdoba, Argentina \\
		}
		\thanks{(Corresponding author: Kanglian Zhao.)}
	}

	\maketitle
	
	\begin{abstract}
		Global navigation satellite systems (GNSSs) can already plan their inter-satellite link (ISL) topology onboard, leaving routing as the last step toward autonomous networking.
		In GNSS ISLs, each satellite links to at most one peer per time slot.
		We exploit this structure to replace end-to-end path search with a single backward sweep over the slots.
		The proposed method, DS-SMFP, propagates earliest-arrival times backward from each destination and outputs a STORE/FORWARD decision for every slot, satellite, and destination, in time linear in the number of slots and with working memory independent of the planning horizon.
		Because each slot needs only its own topology and the delivery state of the next one, routing can steer link selection while the topology is built.
		This slotwise dependency enables Master-Oriented Topology--Routing Co-Construction (MTRCC), which favors traffic to and from a master satellite performing onboard orbit determination.
		In a 24-h, 30-satellite BeiDou scenario, DS-SMFP matches the delays of Contact Graph Routing (CGR) and shortest path tree (SPT) routing while running over 180 times faster and using over 36 times less peak routing working memory.
		Compared with fair contact plan (FCP) topology, MTRCC reduces mean master-related traffic delays by over 10\%, with an approximately 3\% increase in mean PDOP.
		This paper provides a methodological foundation for efficient autonomous GNSS networking.
	\end{abstract}
	

	
	
	
	\section{Introduction}
	\label{sec:introduction}
	
	Inter-satellite links (ISLs) are becoming a key technology for improving
	the autonomy of global navigation satellite systems (GNSSs)~\cite{1,2}. 
	Besides supporting inter-satellite ranging and time synchronization,
	ISLs carry measurement data, telemetry, and navigation information across
	the constellation.
	They therefore provide an important
	infrastructure for autonomous orbit determination (OD) and autonomous
	network operation, reducing the dependence of GNSSs on continuous ground
	support \cite{3,4}. Achieving
	such autonomy, however, requires more than scheduling and establishing
	useful ISLs: the constellation also needs to decide how data packets are forwarded across a time‑varying topology.
	
	The GNSS topology is strongly shaped by its ISL operating mechanism.
	To support diverse ranging requirements, GNSS satellites commonly employ
	a rapidly steerable radio-frequency phased-array terminal for ISL
	establishment~\cite{5}. Its electronic switching capability enables link
	assignments to be organized in time slots: a satellite polls
	different peers over successive slots and communicates with at most one
	scheduled peer in each slot
	\cite{6,7,8taes}. Therefore,
	although several satellites may be geometrically visible at the same time,
	only a small subset of feasible ISLs can actually be activated. Topology
	planning determines which links are scheduled over time so that both
	ranging and communication requirements can be supported
	\cite{9,10}.
	
	Our previous work~\cite{3} has advanced this planning process toward onboard
	autonomy. By exploiting predictable orbital motion, an accurate common
	time reference, and deterministic computation, the fair contact plan
	(FCP) algorithm enables different satellites to independently generate a
	consistent link schedule with low computational demand, without relying on
	a centralized ground planner. This moves network
	control from ground-generated topology planning toward onboard topology
	autonomy. A complete autonomous network, however, must also compute routing
	decisions onboard. If route computation still requires high processing
	cost, it becomes the next bottleneck in autonomous network operation.
	
	Because each GNSS satellite has at most one scheduled peer in a slot, a
	complete contemporaneous end-to-end path is generally unavailable. Data
	delivery is instead formed across multiple slots through store-and-forward
	transmission, making a time-slotted GNSS a representative predictable
	delay-tolerant network (DTN)~\cite{11,21}. Routing in predictable DTNs is commonly formulated as a graph-based
	end-to-end path search problem over known future communication
	opportunities. Contact Graph Routing (CGR)~\cite{12,13} is a representative
	framework of this type: scheduled contacts are represented in a contact
	graph and searched for feasible source--destination routes. Other
	space--time graph formulations adopt different graph
	representations or search structures~\cite{15,19,20,22,23}, but retain the
	same basic paradigm of constructing end-to-end routes over an explicitly
	represented time-varying graph.
	When such general graph-based routing approaches are applied to autonomous
	GNSS operation, three limitations become particularly relevant.

	\paragraph{Sensitivity to planning scale}
	Graph-based end-to-end path search depends on the set of future
	communication opportunities represented over the planning horizon. As
	the horizon becomes longer or the topology changes more frequently, the
	number of time-indexed contacts increases, enlarging the amount of graph
	information that must be stored, organized, and processed. The degree of
	this dependence varies with the underlying graph representation and
	search strategy, and is particularly pronounced in contact-centric
	formulations, where additional contacts directly enlarge the routing
	search space~\cite{12,13}. More compact representations can alleviate
	this growth~\cite{22,23}, but they do not remove the underlying dependence of
	graph-based path search on the number of future communication
	opportunities. In a GNSS, where scheduled ISLs may change every few
	seconds, a full-day schedule therefore presents a nontrivial scalability
	burden for computationally constrained onboard routing.
	
	\paragraph{Insufficient use of GNSS-specific structure}
	General routing methods accommodate a wide range of changing
	networks. GNSS ISLs, however, follow a fixed slot rhythm, and each
	satellite communicates with at most one selected peer in any slot.
	Satellite motion and inter-satellite distances are also highly
	predictable. These characteristics restrict how data can move
	through the network and offer opportunities to simplify the routing
	process, but they are not explicitly used to reorganize the routing computation in general graph-based formulations.
	
	\paragraph{Weak interaction between topology construction and
		routing}
	Topology planning and route computation are strongly related, yet the
	conventional routing formulations largely separate them. Routing-aware FCP~\cite{14} (RA-FCP)
	has used end-to-end routing performance to evaluate a completed contact
	plan and then iteratively refine it. Such a plan-level feedback process
	couples topology and routing only after topology construction and may
	require repeated routing evaluations. Some GNSS studies have
	incorporated telemetry downlink requirements by assuming one-hop
	transmission from ordinary satellites to designated downlink relay
	satellites, or \emph{anchors}, and favoring links between these
	groups~\cite{7,8taes,24}. This couples topology planning with a
	predefined one-hop routing pattern,
	but it does not address unrestricted multi-hop routing between
	arbitrary source--destination pairs.
	
	These limitations lead to two connected questions. 
	First, can the fixed slot rhythm and single-peer link structure of a GNSS be exploited
	to obtain a low-computation routing method suitable for
	onboard execution? Second, can topology construction and route computation influence
	each other directly while the schedule is being formed?
	This paper addresses these two questions through the following contributions.

		1) \textit{Low-complexity onboard routing.}
		A Destination-Seeded Scheduled-Matching Forwarding Propagation
		(DS-SMFP) method is developed to exploit the time-slotted and
		single-peer structure of the GNSS topology. Instead of repeatedly
		searching a graph for complete end-to-end paths, the routing
		computation is organized as slot-to-slot backward state propagation
		and directly generates onboard-executable STORE/FORWARD decisions.
		By retaining only the state needed between adjacent slots, this
		structure yields a predictable $\mathcal{O}(KN^2)$ computational
		complexity, where $K$ is the number of slots and $N$ is the number
		of satellites, while keeping working memory at $\mathcal{O}(N^2)$,
		independent of the planning-horizon length. The method fully
		exploits the matching structure and GNSS-specific orbital
		characteristics to reduce routing computation and working memory.
		
		2) \textit{Topology--routing co-construction mechanism.}
		A Master-Oriented Topology--Routing Co-Construction method (MTRCC) is
		developed to incorporate the routing value of candidate ISLs
		directly into topology construction. Future delivery information guides
		link selection as the schedule is formed, while the ranging-fairness
		objective of FCP is retained. Its weighting mechanism makes the
		trade-off between ranging performance and communication delay
		adjustable, including the relative priorities of traffic toward and
		from a master satellite. This capability is particularly relevant to
		onboard centralized orbit determination, in which ranging data is
		gathered at a master satellite and updated orbit and navigation
		information is disseminated to the constellation.
		
		3) \textit{Evaluation of routing efficiency and
			topology--routing co-construction.}
		The proposed methods are evaluated in a 24 h, 30 satellite BeiDou
		scenario to assess their computational efficiency and communication
		performance. On an identical reverse-order FCP topology, DS-SMFP
		produces the same delay and hop-count
		distributions as Contact Graph Routing (CGR)~\cite{12,13} and shortest-path-tree
		(SPT) routing~\cite{22,23}, while running over
		180 times faster and requiring over 36 times less peak routing
		working memory. The topology experiments compare MTRCC with
		reverse-order FCP and RA-FCP to assess whether
		routing-value link selection can improve communication without
		substantially compromising ranging performance. A representative
		MTRCC configuration reduces mean all-to-all, master-to-all, and
		all-to-master delays from 10.88, 10.89, and 10.91~s under
		reverse-order FCP to 9.93, 9.80, and 9.64~s, respectively, while
		increasing mean PDOP only from 1.0006 to approximately 1.03 and
		maintaining at least nine distinct ranging partners per satellite
		per superframe. Further experiments demonstrate that this
		ranging--communication trade-off can be tuned through the MTRCC
		parameters and that the communication improvements persist for
			master satellites of various orbit types.
		These results demonstrate the lightweight nature of DS-SMFP and the significant advantages of the topology-routing co-construction mechanism in simultaneously maintaining and optimizing topology and routing performance.

	The paper is organized as follows:
	Section II introduces the work related to GNSS autonomy and DTN routing.
	Section~III introduces the GNSS topology and onboard routing model.
	Section~IV develops DS-SMFP and its GNSS-specific refinements.
	Section~V presents topology--routing co-construction and the MTRCC
	instantiation. 
	Section~VI evaluates the proposed methods, and Section~VII
	concludes the paper.
	\section{Related Work}
	\label{sec:related_work}
	
	\subsection{Autonomous GNSS Operation}
	
	Autonomous GNSS operation seeks to maintain essential navigation services
	without continuous support from the ground control segment. Long duration
	autonomy has been studied over periods of 180 days for GPS and 90 days
	for BeiDou~\cite{33,34}. Such operation relies on onboard orbit and clock
	estimation and on inter-satellite networking to exchange measurements
	and distribute updated navigation information.
	
	Studies of autonomous orbit determination (OD) have considered different
	processing architectures. Tang \emph{et al.}~\cite{35} investigated
	centralized OD using BeiDou inter-satellite link (ISL) measurements,
	whereas Guo \emph{et al.}~\cite{36} studied distributed OD.
	Ruan \emph{et al.}~\cite{4} examined orbit and clock estimation from
	raw BDS-3 ISL observations. Cheng \emph{et al.}~\cite{37}
	validated an onboard centralized navigation scheme with attention to
	processor and memory demands, and Xia
	\emph{et al.}~\cite{38} studied long-term distributed OD with reduced
	dependence on uploaded reference orbits.
	
	Autonomy at the network level also requires satellites to coordinate
	ISL establishment. Our work~\cite{3} developed distributed fair
	contact plan design for GNSSs, using predictable orbital motion, a
	shared time reference, and deterministic scheduling to produce
	consistent link assignments onboard. These schedules provide the
	opportunities for exchanging measurement data and navigation updates;
	lightweight onboard routing is another essential component of
	end-to-end autonomous GNSS network operation.
	
	\subsection{Routing in Predictable Delay-Tolerant Networks}
	The scheduled and intermittent connectivity of GNSS ISLs makes them a
	representative predictable DTN, in which
	packets may be stored until later communication opportunities.
	Existing routing studies have primarily addressed general DTN settings
	with known or predictable contacts.
	Jain \emph{et al.}~\cite{19} studied DTN routing under different
	levels of knowledge about future contacts. CGR organizes scheduled contacts as graph vertices and connects them through directed edges when successive transmissions are
	feasible~\cite{12}. Time-aware graph search then identifies routes across
	these contacts~\cite{39}. In a contact multigraph, the vertices
	instead represent network nodes, and time-annotated edges represent
	their scheduled contacts. Shortest-path-tree (SPT) routing uses
	this representation to compute source-rooted routes~\cite{22,23},
	while Moy \emph{et al.}~\cite{13} implemented contact-multigraph routing
	with modified Dijkstra and Yen searches, and Hylton \emph{et al.}~\cite{40}
	analyzed its pathfinding scalability relative to contact-graph routing.
	
	A different representation embeds time directly into the routing graph,
	rather than treating it only as an attribute of scheduled contacts.
	Merugu \emph{et al.}~\cite{15} constructed a space--time graph from
	predictable node mobility and derived forwarding tables indexed by
	packet destination and arrival time,
	Jiang \emph{et al.}~\cite{20} developed space--time-graph
	multipath routing for Earth-observing satellite networks, and
	El Alaoui and Ramamurthy~\cite{41} proposed earliest-arrival
	routing based on a modified temporal graph for interplanetary DTNs.
	These approaches establish a broad foundation for routing over
	scheduled connectivity.
	However, they do not explicitly exploit the GNSS-specific
	single-peer per slot matching structure to organize lightweight
	forwarding computation for autonomous onboard execution.
	\section{GNSS Topology and Onboard Routing Model}
	\label{sec:system_model}
	
    GNSS network has two operational characteristics that directly shape its routing process: the ISL topology changes on a fixed slot basis, and each satellite can communicate with at most one scheduled peer in a slot. The following model formalizes these characteristics and the corresponding information required for onboard data routing.
	
	\subsection{Time-Slotted GNSS Network}
	\label{subsec:time_slotted_network}
	
	Let
	\begin{equation}
		\Vset=\{v_1,v_2,\ldots,v_N\}
	\end{equation}
	denote the set of $N$ GNSS satellites. The scheduling horizon $[t_0,t_K)$ is divided into $K$ equal-length slots indexed by
	\begin{equation}
		\Tset=\{0,1,\ldots,K-1\},
	\end{equation}
	where slot $k\in\Tset$ occupies $[t_k,t_{k+1})$ and
	\begin{equation}
		t_k=t_0+k\Delta t.
	\end{equation}
	Here, $\Delta t$ denotes the slot duration and is the basic time granularity for ISL establishment. Polling ISLs in GNSSs typically employ slots on the order of a few seconds; a 3~s slot is a representative configuration in BeiDou studies~\cite{3,6,7}. 
	
\begin{figure}[t]
		\centering
		\includegraphics[width=0.41\textwidth]{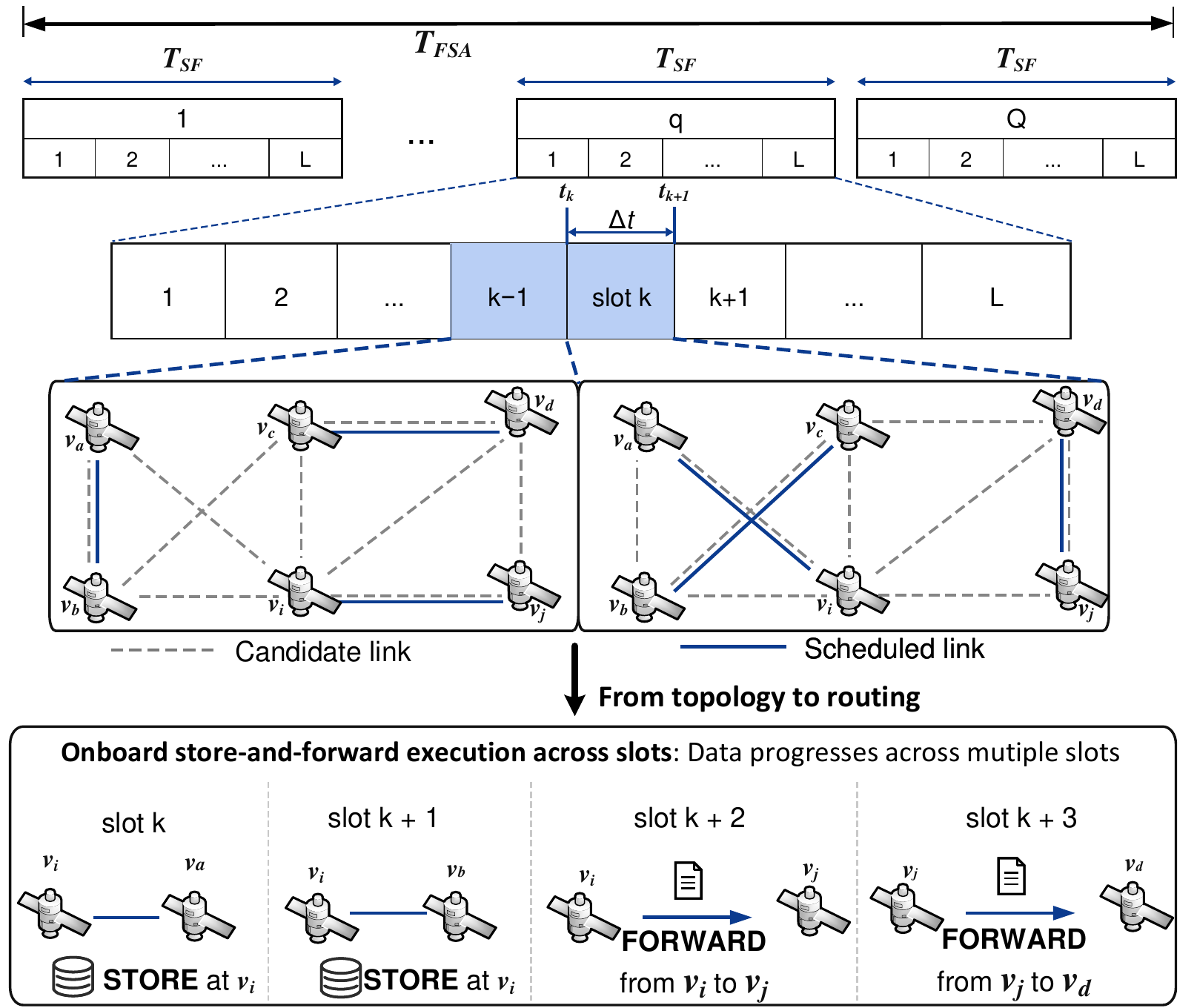}
		\caption{Time-slotted GNSS topology and onboard forwarding model. Satellite visibility is represented by piecewise-static FSA states, while the scheduled ISL topology can change at the slot granularity. Although multiple neighbors can be visible, each satellite has at most one scheduled peer in a slot, resulting in store-and-forward delivery across successive slots.}
		\label{fig:system_model}
	\end{figure}

	The slot-level operation is organized within a hierarchical time structure, as shown in Fig.~\ref{fig:system_model}. Satellite motion continuously changes geometric visibility, and a finite-state-automaton (FSA) representation is used to convert this continuous variation into a sequence of piecewise-static visibility states~\cite{16}. Let $T_{\mathrm{FSA}}$ denote the duration of an FSA state. Two satellites are regarded as mutually visible in an FSA state only if the physical visibility conditions are maintained throughout the entire state; otherwise, that satellite pair is excluded from the candidate links of the state. 
	
	Each FSA state is further divided into $Q$ superframes, and each superframe contains $L$ slots, such that
	\begin{equation}
		T_{\mathrm{FSA}} = Q T_{\mathrm{SF}},
		\qquad
		T_{\mathrm{SF}} = L\Delta t,
	\end{equation}
	where $T_{\mathrm{SF}}$ is the superframe duration. A superframe represents one complete slot-level ISL scheduling cycle under the static visibility of the corresponding FSA state. In the scheduling organization adopted in this work, one $L$ slot scheduling pattern is generated for an FSA state and repeated over its $Q$ superframes. 
	
	\subsection{Visibility and Scheduled ISL Topology}
	\label{subsec:scheduled_topology}
	
	For slot $k$, let
	\begin{equation}
		\mathcal{G}^{\mathrm{vis}}_k
		=
		\bigl(\Vset,\Eset^{\mathrm{vis}}_k\bigr)
	\end{equation}
	denote the candidate visibility graph inherited from the FSA state containing that slot. The edge set $\Eset^{\mathrm{vis}}_k$ contains satellite pairs that remain mutually visible over the corresponding FSA state and therefore satisfy the geometric conditions for ISL scheduling. Several candidate neighbors may be available to the same satellite.
	The actual topology in slot $k$ is obtained by selecting a subset of these candidate links. Let
	\begin{equation}
		\mathcal{M}_k \subseteq \Eset^{\mathrm{vis}}_k
	\end{equation}
	denote the scheduled ISLs. A GNSS satellite is assumed to carry a single ISL terminal and can therefore establish only one ISL in a slot. Accordingly, no two scheduled links in $\mathcal{M}_k$ can share a satellite. In graph-theoretic terminology, $\mathcal{M}_k$ is a \textbf{matching}~\cite{18,17}, i.e.,
	\begin{equation}
		\deg_{\mathcal{M}_k}(v_i)\leq 1,
		\qquad
		\forall v_i\in\Vset,\ \forall k\in\Tset.
	\end{equation}
	
	The satellite paired with $v_i$ in slot $k$ is denoted by $p_k(i)$:
	\begin{equation}
		p_k(i)=
		\begin{cases}
			j, & \text{if } \{v_i,v_j\}\in\mathcal{M}_k,\\
			\varnothing, & \text{if } \deg_{\mathcal{M}_k}(v_i)=0.
		\end{cases}
		\label{eq:scheduled_peer}
	\end{equation}
	Thus, $\Eset^{\mathrm{vis}}_k$ describes all candidate ISLs available to the scheduler, whereas $p_k(i)$ identifies the only peer, if any, that satellite $v_i$ can actually link in slot $k$.

	\subsection{Onboard Store-and-Forward Routing}
	\label{subsec:onboard_routing_requirement}
	
	\subsubsection{High-Rate ISLs and Capacity Margin}
	The central objective of autonomous GNSS operation is to sustain navigation services through inter-satellite ranging, onboard OD, and the dissemination of updated orbit, clock, and navigation information. These operations require the constellation to exchange navigation-related data over its scheduled ISLs. BDS-3 currently employs narrow-beam Ka-band ISLs with a maximum information rate of 200~kb/s~\cite{27}. Beyond supporting inter-satellite ranging, navigation-information dissemination, and telemetry relay, this high-rate ISL infrastructure provides substantial capacity headroom for additional services, including global short-message communication and international search and rescue services \cite{1,28}. Published assessments report a short-message delivery success rate of approximately 98\% with response times generally below 30~s, as well as a success rate of 99.97\% for international search and rescue services~\cite{29,30}. The provision of these services alongside routine navigation operations demonstrates the communication capacity redundant beyond the constellation's essential navigation workload.
	
	\subsubsection{Modest Autonomous-Navigation Data Volumes}
	The navigation-related data are also modest relative to this ISL capability. To obtain an indicative measurement data scale, we examined one day of GNSS observation files from three ground stations in the BeiDou high-precision spatiotemporal service scientific database\footnote{\url{https://www.ntscstdc.cn}}. The files contain approximately 125,410 ground-satellite observation records occupying 4,645~kB, or about 0.037~kB per archived record. Ground-based GNSS observations and inter-satellite measurements have different observation geometries and data formats but both carry compact ranging and timing information; these archived records provide an order-of-magnitude reference for individual measurement data. Existing GNSS studies typically schedule 9--15 ranging links per satellite per minute to obtain diverse observations~\cite{3,6,7}. At the indicative record size, these opportunities correspond to approximately 0.33--0.56~kB per minute, or 44--74~bit/s on average. BeiDou navigation messages are broadcast at approximately 50--1000~bit/s, depending on the signal format~\cite{2}. These data scales are small compared with the high-rate BDS-3 ISL capability.
	
	\subsubsection{Capacity-Sufficient Store-and-Forward Routing}
	Given the ample ISL capacity available for these essential exchanges, we consider a capacity-sufficient operating regime in which navigation-related packets complete transmission over their scheduled ISLs within the assigned slots and experience almost no queueing delay. End-to-end delivery time is thus governed by the sequence of scheduled communication opportunities and the waiting intervals between them. Because each satellite has at most one scheduled peer in a slot, a packet can traverse at most one distinct ISL in that slot; an intermediate relay continues forwarding in subsequent slots.
	
	For each slot $k$, every satellite $v_i$ requires a routing decision for
	packets destined for any other satellite $v_d$ ($i\neq d$). A packet held by
	$v_i$ at the beginning of slot $k$ can either remain onboard until the next
	slot or be transmitted to its scheduled peer $v_{p_k(i)}$. Denote the
	corresponding decision by
	\begin{equation}
		a_k(i,d)\in\{\mathrm{STORE},\mathrm{FORWARD}\}.
		\label{eq:sm_action}
	\end{equation}
	Here, \textnormal{STORE} retains the packet at $v_i$, whereas
	\textnormal{FORWARD} sends it through the current scheduled ISL. When
	$p_k(i)=\varnothing$, the applicable action is \textnormal{STORE}.
	
	With the scheduled links change across slots, successive routing decisions
	form an end-to-end store-and-forward path: each satellite carrying the packet
	uses its decision for the current slot and destination, as shown in Fig.~\ref{fig:system_model}. The onboard routing
	requirement is to determine the actions $a_k(i,d)$ for all relevant
	slot--satellite--destination states so that the resulting paths achieve the
	earliest destination arrival times, with the fewest ISL hops among
	paths having the same arrival time.
	
	Accordingly, an onboard entry in the local forwarding table of satellite $v_i$ takes
	the form
	\begin{equation}
		\left\langle k,\,d,\,p_k(i),\,a_k(i,d)\right\rangle,
		\label{eq:sm_forwarding_entry}
	\end{equation}
	where $k$ and $d$ identify the slot and destination, $p_k(i)$ is the peer
	specified by the scheduled topology, and $a_k(i,d)$ specifies the action to
	execute. The satellite identity $i$ is implicit in its local table. The next
	section develops DS-SMFP to generate these forwarding entries over the
	scheduled matching sequence.
	
	\section{Low-Complexity Onboard Routing}
	\label{sec:low_complexity_routing}
	
	The time-indexed forwarding actions defined in
	Section~\ref{subsec:onboard_routing_requirement} can be generated by
	propagating future delivery state from later slots toward earlier
	ones. This section develops the
	\emph{Destination-Seeded Scheduled-Matching Forwarding Propagation}
	(DS-SMFP) method and shows how the resulting actions are compiled directly
	into onboard routing entries.
	
	\subsection{Destination-Seeded Future-Delivery States and Terminal Initialization}
	\label{subsec:future_delivery_state}
	
	For a packet held at satellite $v_i$ at the beginning of slot $k$ and destined for $v_d$, let $J_k(i,d)$ denote its earliest achievable arrival time at $v_d$ under the remaining scheduled topology. Let $h_k(i,d)$ denote the fewest ISL forwarding hops among routes attaining that arrival time. Thus, $(J_k(i,d),h_k(i,d))$ characterizes the optimal future delivery, with arrival time as the primary criterion and hop count as the secondary criterion.
	
	The destination states initialize the backward propagation. At the end of the planning horizon, $t_K$,
	\begin{align}
		J_K(i,d)&=\begin{cases}
			t_K,&i=d,\\
			+\infty,&i\ne d,
		\end{cases}
		\label{eq:terminal_arrival}\\
		h_K(i,d)&=\begin{cases}
			0,&i=d,\\
			+\infty,&i\ne d.
		\end{cases}
		\label{eq:terminal_hops}
	\end{align}
	At an earlier slot, a packet already at its destination satisfies
	\begin{equation}
		J_k(d,d)=t_k,\qquad h_k(d,d)=0.
		\label{eq:destination_absorbing}
	\end{equation}
	These boundary conditions seed the delivery-state propagation toward preceding slots.
	
	\subsection{Slot-by-Slot Backward Forwarding Computation}
	\label{subsec:slot_propagation}
	
	DS-SMFP processes the scheduled slots in reverse order, $k=K-1,K-2,\ldots,0$. When slot $k$ is processed, all future-delivery states at slot $k+1$ are available. For $i\ne d$, retaining the packet at $v_i$ gives the \textnormal{STORE} candidate
	\begin{equation}
		\begin{aligned}
			J_k^{\mathrm S}(i,d)&=J_{k+1}(i,d),\\
			h_k^{\mathrm S}(i,d)&=h_{k+1}(i,d).
		\end{aligned}
		\label{eq:store_candidate}
	\end{equation}
	When $p_k(i)=j$, transmitting the packet to $v_j$ gives the \textnormal{FORWARD} candidate
	\begin{align}
		J_k^{\mathrm F}(i,d)&=\begin{cases}
			t_k+\tau_k(i,d),&j=d,\\
			J_{k+1}(j,d),&j\ne d,
		\end{cases}
		\label{eq:forward_arrival}\\
		h_k^{\mathrm F}(i,d)&=\begin{cases}
			1,&j=d,\\
			h_{k+1}(j,d)+1,&j\ne d,
		\end{cases}
		\label{eq:forward_hops}
	\end{align}
	where $\tau_k(i,j)$ is the one-way propagation time of the scheduled ISL. Direct transmission to $v_d$ completes delivery in slot $k$. For an intermediate relay $v_j$, the matching makes $v_i$ its only scheduled peer during slot $k$. A subsequent transmission from $v_j$ to a distinct satellite can therefore occur only in a later slot, making its future state at $k+1$ the exact continuation.
	
	For two delivery outcomes, define the lexicographic preference
	\begin{equation}
		(x,r)\prec(y,s)
		\iff (x<y)\ \text{or}\ (x=y\ \text{and}\ r<s).
		\label{eq:lexicographic_order}
	\end{equation}
	The optimal delivery state is
	\begin{equation}
		(J_k(i,d),h_k(i,d))\\[-0.5mm]
		=\operatorname*{lexmin}\!\left\{
		(J_k^{\mathrm S},h_k^{\mathrm S}),
		(J_k^{\mathrm F},h_k^{\mathrm F})\right\},
		\label{eq:selected_state}
	\end{equation}
	where the arguments $(i,d)$ on the right are omitted for compactness. The FORWARD candidate applies when a scheduled peer exists. The preferred candidate directly determines $a_k(i,d)$; an exact tie selects STORE. 
	An unreachable destination retains an infinite delivery-state label.
	
	\noindent\textbf{Proposition 1:}
	For any given sequence of scheduled matchings under the slot-level store-and-forward model, \eqref{eq:terminal_arrival}--\eqref{eq:selected_state} yield the earliest reachable arrival time for every state $(k,i,d)$ and the fewest forwarding hops among routes attaining that arrival time.
	
	\noindent\emph{Proof:}
	The terminal states at $k=K$ are exact. Assume that all delivery states at slot $k+1$ are optimal. For $i=d$, \eqref{eq:destination_absorbing} gives immediate delivery. For $i\ne d$, the matching supplies at most one peer $v_j$. STORE retains the packet at $v_i$, giving the exact continuation $(J_{k+1}(i,d),h_{k+1}(i,d))$. FORWARD delivers the packet to $v_j$ during slot $k$. When $j=d$, delivery finishes at $t_k+\tau_k(i,d)$ in one hop. When $j\ne d$, $v_j$ is matched solely with $v_i$ during slot $k$ and can forward toward any distinct next relay only from slot $k+1$ onward. Its optimal continuation is therefore precisely $(J_{k+1}(j,d),h_{k+1}(j,d))$. These candidates exhaust the feasible first actions and their optimal continuations. Their lexicographic minimum is optimal at slot $k$, completing the backward induction. \hfill$\square$
	
	The recurrence applies to every satellite--destination state. Define the slot-level arrival-time and hop-count matrices
	\begin{equation}
		\mathbf J_k=[J_k(i,d)]_{N\times N},\qquad
		\mathbf H_k=[h_k(i,d)]_{N\times N}.
		\label{eq:state_matrices}
	\end{equation}
	For satellite $v_i$, row $i$ of $(\mathbf J_{k+1},\mathbf H_{k+1})$ supplies STORE outcomes for all destinations, while the scheduled peer's row supplies FORWARD continuations. Comparing the rows yields the current states and forwarding actions for $v_i$. Applying the update to every satellite produces
	\begin{equation}
		\mathbf J_K\longrightarrow\mathbf J_{K-1}
		\longrightarrow\cdots\longrightarrow\mathbf J_0,
		\label{eq:backward_matrix_sequence}
	\end{equation}
	with the hop-count matrices propagated in the same order.
	
	Such a direct \emph{satellite-wise implementation} evaluates the STORE/FORWARD decision independently for each ordered satellite--destination pair $(i,d)$, $i\ne d$, requiring $N(N-1)$ state evaluations per slot. It reads the next-slot matrices $(\mathbf J_{k+1},\mathbf H_{k+1})$ and writes the results into $(\mathbf J_k,\mathbf H_k)$, maintaining two adjacent time-layer matrices in working memory.
	
	\subsection{Matching-Based Computation and Memory Reduction}
	\label{subsec:pair_inplace_propagation}
	
	\subsubsection{Matching-Wise Joint Evaluation}
	Each scheduled link $\{v_i,v_j\}\in\mathcal M_k$ connects two satellites whose forwarding decisions for a common destination depend on the same pair of future states. This permits a \emph{matching-wise implementation} that evaluates both endpoints together rather than performing two satellite-wise comparisons.
	
	For a destination $d\notin\{i,j\}$, denote the two future states by
	\begin{equation}
		\begin{aligned}
			X&=(J_{k+1}(i,d),h_{k+1}(i,d)),\\
			Y&=(J_{k+1}(j,d),h_{k+1}(j,d)).
		\end{aligned}
		\label{eq:pair_future_states}
	\end{equation}
	Write $X=(X_1,X_2)$ and $Y=(Y_1,Y_2)$, where the first component is arrival time and the second is hop count. The STORE outcome for $v_i$ is $X$, and its FORWARD outcome through $v_j$ is $(Y_1,Y_2+1)$. The corresponding outcomes for $v_j$ are $Y$ and $(X_1,X_2+1)$. Hence,
	\begin{align}
		a_k(i,d)=\mathrm{FORWARD}
		&\iff (Y_1,Y_2+1)\prec X,
		\label{eq:pair_forward_i}\\
		a_k(j,d)=\mathrm{FORWARD}
		&\iff (X_1,X_2+1)\prec Y.
		\label{eq:pair_forward_j}
	\end{align}
	
	These two decisions share the same arrival-time ordering. When $X_1<Y_1$, $v_j$ forwards toward $v_i$, while $v_i$ stores; when $Y_1<X_1$, the actions are reversed. For equal finite arrival times, forwarding from one endpoint is beneficial only when the other endpoint offers at least two fewer remaining hops. Otherwise, both endpoints store. When both future destinations are unreachable, both also store. Accordingly, each pair--destination evaluation yields
	\begin{equation}
		(a_k(i,d),a_k(j,d))\in\{(\mathrm F,\mathrm S),(\mathrm S,\mathrm F),
		(\mathrm S,\mathrm S)\},
		\label{eq:pair_exclusive_actions}
	\end{equation}
	where $\mathrm F$ and $\mathrm S$ abbreviate FORWARD and STORE. The endpoint destinations $d=i$ and $d=j$ are evaluated by the direct-delivery rule in \eqref{eq:forward_arrival}--\eqref{eq:forward_hops}.
	
	The matching-wise implementation therefore uses a \emph{matching-pair--destination} comparison as its basic unit. Each of the $|\mathcal M_k|$ matched pairs has $N-2$ third-party destinations. Satellite-wise processing performs $2|\mathcal M_k|(N-2)$ separate directional comparisons for these states, whereas matching-wise processing uses $|\mathcal M_k|(N-2)$ shared pair evaluations to determine both actions. 
	The shared comparison reduces repeated state evaluations while retaining both endpoint forwarding decisions.
	
	\subsubsection{Matching-Wise In-Place Delivery State Propagation}
	The independence of matching pairs also allows current-slot delivery states to overwrite the next-slot states in the same matrices. Let the working matrices $\mathbf J$ and $\mathbf H$ initially hold $(\mathbf J_{k+1},\mathbf H_{k+1})$. For each matching edge and third-party destination, the original states $X$ and $Y$ in \eqref{eq:pair_future_states} are read into local temporaries before either entry is updated. The two results are then written back as
	\begin{equation}
		\begin{aligned}
			(J(i,d),h(i,d))&\leftarrow
			\operatorname*{lexmin}\{X,(Y_1,Y_2+1)\},\\
			(J(j,d),h(j,d))&\leftarrow
			\operatorname*{lexmin}\{Y,(X_1,X_2+1)\}.
		\end{aligned}
		\label{eq:inplace_pair_update}
	\end{equation}
	Both assignments use the captured next-slot states $X$ and $Y$. Each matching pair updates only its two satellite rows, while all other pairs occupy disjoint rows and retain their next-slot values until processed. Within a pair, each destination column uses its own captured state values. Consequently, neither updates within a pair nor updates across different pairs alter the future states needed by subsequent evaluations. Direct-delivery entries follow \eqref{eq:forward_arrival}--\eqref{eq:forward_hops}; unmatched rows retain their states, and diagonal entries follow \eqref{eq:destination_absorbing}. Once all pairs have been processed, the same $\mathbf J$ and $\mathbf H$ represent $(\mathbf J_k,\mathbf H_k)$.
	
	Matching-wise joint evaluation reduces repeated delivery-state comparisons, while matching-wise in-place propagation replaces two adjacent time layers with one resident arrival-time matrix and one resident hop-count matrix. Both improvements follow from the disjoint pair structure of the scheduled ISLs.
	
	\subsection{GNSS-Orbit-Enabled Integer Delivery-Slot Labels}
	\label{subsec:integer_slot_labels}
	
	The GNSS orbital geometry permits a further reduction by replacing floating-point arrival times with integer delivery-slot labels. Reported BeiDou ISL distances extend to approximately $70{,}000$~km, corresponding to a one-way propagation time below $0.24$~s~\cite{27}. For the $3$ s slots considered here, each scheduled ISL satisfies $0\le\tau_k(i,j)<\Delta t$. A final ISL transmission in slot $r$ consequently yields a nominal destination arrival within $[t_r,t_{r+1})$.
	
	An earlier final-delivery slot always gives an earlier arrival. For a fixed destination and final-delivery slot, the scheduled matching also specifies a unique incoming link, and hence an identical final-hop propagation time. Thus, the final-delivery slot preserves the earliest-arrival ordering exactly, with hop count distinguishing routes having equal arrival times.
	
	Define $R_k(i,d)$, for $i\ne d$, as the slot index of the final ISL transmission on an earliest-arrival route from $v_i$ to $v_d$ starting at slot $k$. Unreachable states carry $+\infty$. For the destination itself, set $R_k(d,d)=k$; at the terminal boundary, $R_K(d,d)=K$ and $R_K(i,d)=+\infty$ for $i\ne d$. The floating-point state matrix $\mathbf J$ is thereby replaced by the integer matrix $\mathbf R=[R(i,d)]_{N\times N}$, while $\mathbf H$ retains hop counts.
	
	The STORE and FORWARD candidates become
	\begin{align}
		R_k^{\mathrm S}(i,d)&=R_{k+1}(i,d),
		\label{eq:slot_store}\\
		R_k^{\mathrm F}(i,d)&=\begin{cases}
			k,&p_k(i)=d,\\
			R_{k+1}(p_k(i),d),&p_k(i)\ne d.
		\end{cases}
		\label{eq:slot_forward}
	\end{align}
	Their hop counts follow \eqref{eq:store_candidate} and \eqref{eq:forward_hops}. Comparing the $(R,h)$ candidates lexicographically therefore produces the same forwarding decisions as comparing $(J,h)$.
	
	When a nominal absolute arrival time is needed, the delivery-slot label also identifies the final incoming link. For a reachable state with $i\ne d$, let $r=R_k(i,d)$; then
	\begin{equation}
		J_k(i,d)=t_r+\tau_r(p_r(d),d),\qquad i\ne d.
		\label{eq:arrival_from_slot_label}
	\end{equation}

	The reduced label width also lowers the memory traffic of backward propagation. A $24$ h horizon with $3$ s slots contains $28{,}800$ slots, which fit in an unsigned 16-bit integer field. Using 16-bit unsigned integers for both the delivery-slot matrix $\mathbf R$ and hop-count matrix $\mathbf H$, the matching-wise in-place implementation requires $(2+2)N^2=4N^2$ bytes of core state storage. For comparison, the satellite-wise implementation uses 64-bit floating-point arrival-time matrix $\mathbf J$ and 16-bit hop-count matrix $\mathbf H$ in two time-layer matrices, requiring $2(8+2)N^2=20N^2$ bytes. Matching-wise in-place propagation with floating-point arrival-time matrix $\mathbf J$ reduces this to $(8+2)N^2=10N^2$ bytes, while integer delivery-slot labels further reduce the core state storage to $4N^2$ bytes.

	\subsection{Deterministic Complexity and Direct Onboard Routing-Table Generation}
	\label{subsec:all_state_computation}
	
	Algorithm~\ref{alg:dssmfp} combines backward slot propagation, matching-wise joint evaluation, in-place state updates, and integer delivery-slot labels. Each slot produces the preceding-slot delivery states and the forwarding actions for every satellite and destination.
	
	\begin{algorithm}[t]
		\caption{DS-SMFP Slot-by-Slot Forwarding Compilation}
		\label{alg:dssmfp}
		\KwIn{Scheduled matchings $\mathcal M_k$, $k=0,\ldots,K-1$}
		\KwOut{Forwarding flags $B_k(i,d)$ for all $k,i,d$}
		Initialize integer matrices $\mathbf R,\mathbf H$ at slot $K$\;
		\For{$k=K-1,K-2,\ldots,0$}{
			Initialize all entries of $B_k$ as STORE\;
			\ForEach{$\{v_i,v_j\}\in\mathcal M_k$}{
				\For{$d\notin\{i,j\}$}{
					Read and retain both future states\;
					Jointly evaluate and update both states in place\;
					Record the two forwarding flags\;
				}
				Evaluate direct delivery for $d=i$ and $d=j$\;
			}
			Set $R_k(d,d)=k$ and $h_k(d,d)=0$ for all $d$\;
			Emit $B_k$\;
		}
	\end{algorithm}

	Each slot has at most $\lfloor N/2\rfloor$ matching edges, and each edge is jointly evaluated for at most $N$ destinations. The number of matching-pair evaluations per slot is therefore bounded by $\lfloor N/2\rfloor N$, yielding $\mathcal O(N^2)$ complexity. Initializing and emitting the complete $N\times N$ forwarding-policy array also requires $\mathcal O(N^2)$ work per slot. The total computational complexity over $K$ slots is therefore
	\begin{equation}
		\mathcal O(KN^2),
		\label{eq:time_complexity}
	\end{equation}
With in-place propagation, the algorithm maintains only the two
$N\times N$ working matrices $\mathbf R$ and $\mathbf H$.
These arrays require $\mathcal O(N^2)$ storage, giving a rolling
working memory complexity of
\begin{equation}
	\mathcal O(N^2).
	\label{eq:working_memory}
\end{equation}
	The working memory is independent of the planning-horizon length. 
	
	Each computed action is recorded in a binary forwarding flag:
	\begin{equation}
		B_k(i,d)=\begin{cases}
			1,&a_k(i,d)=\mathrm{FORWARD},\\
			0,&a_k(i,d)=\mathrm{STORE}.
		\end{cases}
		\label{eq:policy_encoding}
	\end{equation}
	When the destination is unreachable within the planning horizon, the internal delivery label remains infinite and the executable forwarding flag is STORE. Together with the scheduled peer $p_k(i)$, $B_k(i,d)$ directly supplies the local forwarding-table entry $\langle k,d,p_k(i),a_k(i,d)\rangle$ in \eqref{eq:sm_forwarding_entry}. Satellite $v_i$ retains its own $\mathcal O(KN)$ entries, while the complete constellation-wide policy contains $\mathcal O(KN^2)$ entries. Thus, DS-SMFP produces the onboard forwarding table itself as the routing result. In path-oriented routing, an end-to-end route is typically first identified and then translated into per-node forwarding entries; here, each optimal action is directly written to its corresponding table entry during the backward sweep. 
	
	\subsection{Relationship to Existing Backward Recursive Routing}
	\subsubsection{Common Principle of Backward Recursion}
	The backward propagation in DS-SMFP builds on an established idea in
	routing over predictable temporal networks: forwarding decisions at an
	earlier time can be derived from delivery opportunities available later.
	Merugu \emph{et al.}~\cite{15} represented predictable connectivity in
	space and time. Their method initializes time-dependent delays through a
	backward procedure and then applies an intermediate-node recurrence to
	combine multi-hop routes, with a reported route-computation complexity of
	$\mathcal{O}(N^3T)$ for $N$ nodes and $T$ time steps. In public-transit
	routing, the profile variant of the Connection Scan Algorithm
	(CSA)~\cite{32} scans scheduled connections in reverse
	departure-time order and reuses downstream journey profiles. In its basic
	implementation, a reverse scan for one destination processes $C$
	connection events and queries profiles containing at most $P$ retained
	departure--arrival time records.
	With presorted connections, binary-search
	queries, and amortized constant-time profile updates, separate scans for
	all $N$ destinations require $\mathcal{O}(NC\log P)$ time.
	\subsubsection{Deterministic Slotwise Recursion under Matchings}
	
	DS-SMFP realizes this idea through the matching structure
	of GNSS ISLs. In each slot $k$, a satellite has at most one scheduled peer. 
	A subsequent hop toward a different satellite must
	therefore begin in a later slot. Accordingly, the STORE and unique
	FORWARD actions admit exact continuations from the states at slot
	$k+1$. This yields a direct
	slot-by-slot recurrence for every satellite--destination pair, requiring
	$\mathcal{O}(N^2)$ work per slot and $\mathcal{O}(KN^2)$ over $K$ slots.
	The fixed per-slot computational bound and regular state accesses are
	particularly valuable for resource-constrained onboard processors.
	
	\subsubsection{Further GNSS-Specific Refinements}
	As detailed previously, the combination of GNSS orbits and matching structure further reduces DS-SMFP computation and memory beyond the basic backward recurrence. 
	For each matched pair and destination, a shared comparison determines both directional forwarding decisions, halving logical evaluations. Disjoint matching endpoints allow in-place state matrix updates, eliminating one full time-layer buffer. Bounded ISL propagation times and unique incoming links enable integer final-delivery-slot labels. These improvements reduce core state-array storage while directly producing the slot-indexed STORE/FORWARD table.
	More importantly, by leveraging the slot level recurrence of DS-SMFP, we establish a tightly coupled construction process for routing and topology, as developed in
	Section~\ref{sec:topology_forwarding_integration}.
	
	\section{Topology--Routing Integration}
	\label{sec:topology_forwarding_integration}
	
	The scheduled topology determines the forwarding opportunities available
	to each satellite, while routing performance reflects the communication
	value of these opportunities. The slot-level dependency of DS-SMFP allows
	the value to be integrated as the topology is constructed in
	reverse time order.
	
	\subsection{Construction-Stage Topology--Routing Co-Construction}
	\label{subsec:construction_stage_coconstruction}
	
	A conventional topology--routing workflow first generates the complete
	ISL schedule and then computes routing decisions over that schedule.
	The local recursion of DS-SMFP removes this complete-plan dependency:
	as topology matchings are generated in reverse time order, each completed
	$\mathcal{M}_k$ can be processed immediately using the future-delivery
	states at slot $k+1$,
	\begin{equation}
		(\mathcal M_k,\mathbf R_{k+1},\mathbf H_{k+1})
		\longrightarrow
		(\mathbf R_k,\mathbf H_k,\mathbf B_k).
		\label{eq:slot_compilation_dependency}
	\end{equation}
	When matching selection is independent of routing, this slotwise dependency
	supports streaming topology--routing computation: the topology module
	produces matchings in reverse time order, while the routing module
	processes each matching as it becomes available. On separate computing
	resources, their execution can overlap, thereby reducing the overall computation time. 
	In this arrangement, information
	flows only from topology to routing: the matching is selected before its forwarding decisions are computed.
	
	A closer coupling becomes possible because $\mathbf R_{k+1}$
	already summarizes the downstream delivery opportunities when slot $k$
	is being constructed. These states can therefore be used to evaluate the
	routing benefit of candidate ISLs before the current matching is selected.
	
	For a candidate edge $e\in\Eset^{\mathrm{vis}}_k$, let
	$\widehat{G}_k(e)$ denote its normalized routing value derived from
	$\mathbf{R}_{k+1}$. The specific mapping from future-delivery states to
	$\widehat{G}_k(e)$ depends on the importance assigned to different
	source--destination tasks. The topology constructor combines this routing
	value with the original topology objective to determine the scheduled
	matching $\mathcal{M}_k$. Once $\mathcal{M}_k$ is fixed, DS-SMFP
	immediately produces $\mathbf{R}_k$, $\mathbf{H}_k$, and the forwarding
	policy for slot $k$. The newly obtained future-delivery states then
	participate in constructing the preceding slot. The resulting dependency is
	\begin{equation}
		\mathbf{R}_{k+1}
		\longrightarrow
		\widehat{G}_k
		\longrightarrow
		\mathcal{M}_k
		\longrightarrow
		\mathbf{R}_k
		\longrightarrow
		\widehat{G}_{k-1}
		\longrightarrow
		\mathcal{M}_{k-1}.
		\label{eq:coconstruction_chain}
	\end{equation}
	This relation changes routing from a post-topology computation into a
	construction-stage information source. The scheduled topology determines
	the current routing action, while the forwarding states obtained from
	future slots influence which links are scheduled next on the reverse time
	axis. The same mechanism can therefore support different mission-oriented
	topology biases by changing how source--destination delivery improvements
	are mapped into $\widehat{G}_k(e)$, without changing the DS-SMFP state
	propagation itself.

	\subsection{For Autonomous Orbit Determination: A Master-Satellite-Oriented Topology--Routing Co-Construction (MTRCC) Case}
	\label{subsec:master_satellite_case}
	
\begin{figure}[t]
		\centering
		\includegraphics[width=0.41\textwidth]{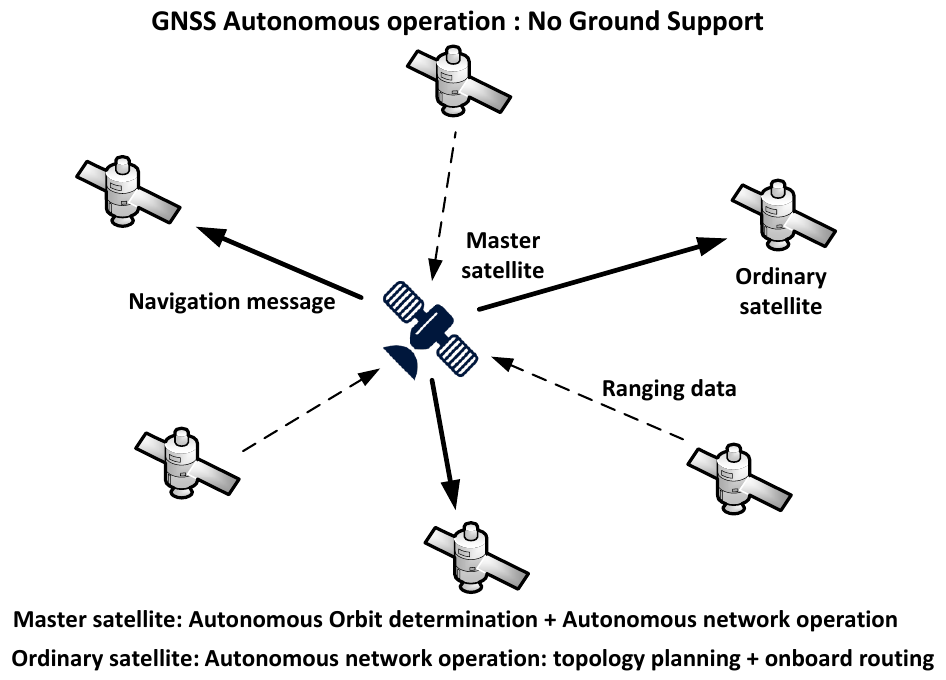}
		\caption{Bidirectional master-satellite traffic in autonomous GNSS operation. Measurement and state information is aggregated toward the master for onboard orbit determination, while updated orbit and navigation information is subsequently disseminated throughout the constellation.}
		\label{fig:master_od}
	\end{figure}

	A representative application is autonomous GNSS orbit determination (OD), illustrated in Fig.~\ref{fig:master_od}. A satellite with comparatively strong onboard processing capability is designated as the \textbf{master satellite}, $v_m$. Measurement information generated throughout the constellation is forwarded toward $v_m$ for onboard orbit determination; updated orbit and navigation information is subsequently distributed from $v_m$ to the other satellites~\cite{35,3}. The master therefore
	becomes a communication-critical node in the autonomous operation loop.
	To reflect the importance of these two traffic directions, define source and destination weights
	\begin{equation}
		w_{\mathrm{src}}(s)
		=\begin{cases}
			W_{\mathrm M}^{\mathrm{src}},s=m,\\
			W_{\mathrm O}^{\mathrm{src}},s\ne m,
		\end{cases}
		\label{eq:master_source_destination_weight}\\
		w_{\mathrm{dst}}(d)
		=\begin{cases}
			W_{\mathrm M}^{\mathrm{dst}},d=m,\\
			W_{\mathrm O}^{\mathrm{dst}},d\ne m,
		\end{cases}
	\end{equation}
	where the subscripts $\mathrm M$ and $\mathrm O$ refer to the master and ordinary satellites. A source--destination task $(s,d)$, $s\ne d$, receives weight
	\begin{equation}
		w_{\mathrm{OD}}(s,d)=w_{\mathrm{src}}(s)w_{\mathrm{dst}}(d).
		\label{eq:od_importance_weight}
	\end{equation}
	Hence, master-to-other tasks are strengthened through the
	source weight, other-to-master tasks are strengthened through
	the destination weight, and ordinary other-to-other tasks retain
	their background importance.
	
	Consider a candidate link $e=\{v_i,v_j\}\in\Eset_k^{\mathrm{vis}}$. Retaining a packet destined for $v_d$ at $v_i$ preserves the future delivery-slot label $R_{k+1}(i,d)$. If the candidate link is selected and the packet is forwarded to $v_j$, its prospective final-delivery label is
	\begin{equation}
		U_k(i\!\rightarrow\!j,d)=
		\begin{cases}
			k,&d=j,\\
			R_{k+1}(j,d),&d\ne j.
		\end{cases}
		\label{eq:edge_use_arrival}
	\end{equation}
	The reverse direction $j\rightarrow i$ is evaluated symmetrically. 
	
	To evaluate candidate-link improvements on a common time scale, let $T_{\mathrm c}>0$ be a bound on the remaining delivery cost. A delivery-slot label $r$ is converted into a bounded delay from slot $k$ as
	\begin{equation}
		C_k(r)=
		\begin{cases}
			\min\!\bigl\{T_{\mathrm c},
			\Delta t\max(0,r-k)\bigr\},&r<+\infty,\\
			T_{\mathrm c},&r=+\infty.
		\end{cases}
		\label{eq:bounded_route_cost}
	\end{equation}
	Here, $\Delta t$ converts slot differences into seconds. The directed benefit of using $v_i\rightarrow v_j$ for destination $v_d$ is
	\begin{equation}
		\begin{aligned}
			g_k(i\!\rightarrow\!j,d)=\max\bigl\{0,\,
			&C_k\!\left(R_{k+1}(i,d)\right)\\[-0.5mm]
			&-C_k\!\left(U_k(i\!\rightarrow\!j,d)\right)
			\bigr\}.
		\end{aligned}
		\label{eq:directional_route_gain}
	\end{equation}
	Thus, the routing gain measures the reduction in the bounded remaining delivery delay produced by scheduling the candidate link. 
	The routing value of candidate link $e=\{v_i,v_j\}$ aggregates the gains of both directed uses across destinations:
	\begin{equation}
		\begin{aligned}
			G_k(e)={}&\frac{w_{\mathrm{src}}(i)}{N-1}
			\sum_{\substack{d=1\\d\ne i}}^{N}
			w_{\mathrm{dst}}(d)\,g_k(i\!\rightarrow\!j,d)\\
			&+\frac{w_{\mathrm{src}}(j)}{N-1}
			\sum_{\substack{d=1\\d\ne j}}^{N}
			w_{\mathrm{dst}}(d)\,g_k(j\!\rightarrow\!i,d).
		\end{aligned}
		\label{eq:raw_route_value}
	\end{equation}
	The two terms measure, respectively, how the candidate link improves delivery from $v_i$ through $v_j$ and from $v_j$ through $v_i$. The future-delivery labels encode downstream multi-hop opportunities, allowing links between ordinary satellites to contribute to master-related paths.
	
	Let $F_k(e)\ge0$ denote the FCP fairness score of candidate link $e$~\cite{11}. Define the maxima of the candidate-link scores as
	\begin{equation}
		\begin{aligned}
			F_k^{\max}&=\max_{e'\in\Eset_k^{\mathrm{vis}}}F_k(e'),\\
			G_k^{\max}&=\max_{e'\in\Eset_k^{\mathrm{vis}}}G_k(e').
		\end{aligned}
		\label{eq:score_maxima}
	\end{equation}
	The normalized fairness and routing scores over the
	candidate-edge set are
	\begin{align}
		\widehat F_k(e)&=
		\begin{cases}
			F_k(e)/F_k^{\max},&F_k^{\max}>0,\\
			0,&\text{otherwise},
		\end{cases}
		\label{eq:fairness_normalization}\\
		\widehat G_k(e)&=
		\begin{cases}
			G_k(e)/G_k^{\max},&G_k^{\max}>0,\\
			0,&\text{otherwise}.
		\end{cases}
		\label{eq:routing_normalization}
	\end{align}
	The joint candidate-edge weight is
	\begin{equation}
		W_k(e)=\widehat F_k(e)+\lambda\widehat G_k(e),
		\qquad\lambda\ge0,
		\label{eq:joint_edge_weight}
	\end{equation}
	where $\lambda$ sets the relative strength of routing feedback. Let $\mathfrak M_k$ denote the set of feasible matchings over $\Eset_k^{\mathrm{vis}}$. The topology scheduled in slot $k$ is obtained by
	\begin{equation}
		\mathcal M_k=\underset{\mathcal M\in\mathfrak M_k}{\arg\max}
		\sum_{e\in\mathcal M}W_k(e),
		\label{eq:joint_matching}
	\end{equation}
	which is a maximum-weight matching problem~\cite{25,7}. The resulting $\mathcal M_k$ is immediately processed by DS-SMFP to produce $\mathbf R_k$, $\mathbf H_k$, and the forwarding flags for slot $k$. Setting $\lambda=0$ removes the routing
	term and recovers the baseline reverse-order FCP topology construction.
	
	For balanced master-to-all and all-to-master priorities, the weights may be set to
	\begin{equation}
		W_{\mathrm M}^{\mathrm{src}}=W_{\mathrm M}^{\mathrm{dst}}=\rho,~
		W_{\mathrm O}^{\mathrm{src}}=W_{\mathrm O}^{\mathrm{dst}}=1,
		\qquad\rho\ge1.
		\label{eq:symmetric_master_weight}
	\end{equation}
	Both master-related directions then receive weight $\rho$, while ordinary source--destination traffic retains unit weight. Separate source and destination weights accommodate different aggregation and dissemination priorities.
	For convenience, we hereafter refer to this method as Master-Oriented Topology--Routing Co-Construction (\textbf{MTRCC}).
	
	For each candidate edge, \eqref{eq:raw_route_value} requires at most $\mathcal O(N)$ delivery-gain evaluations. Across the candidate visibility graph, the routing-value computation therefore requires $\mathcal O(|\Eset_k^{\mathrm{vis}}|N)$ operations per topology phase. 

	\section{Evaluation}
	This section evaluates the lightweight characteristics of DS-SMFP for onboard routing, and the ranging--communication performance achieved by MTRCC.

	\subsection{Experimental Setup}
	\label{subsec:experimental_setup}
		\begin{table}[t]
		\centering
		\caption{Common Simulation Parameters}
		\label{tab:experimental_parameters}
		\footnotesize
		\begin{tabular}{@{}p{0.47\columnwidth}p{0.45\columnwidth}@{}}
			\toprule
			\textbf{Parameter} & \textbf{Setting} \\
			\midrule
			Planning horizon & 24~h \\
			Constellation & 24 MEO, 3 IGSO, 3 GEO \\
			MEO configuration & Walker-$\delta$ $24/3/1$ \\
			MEO altitude / inclination & 21,528~km / $55^{\circ}$ \\
			IGSO altitude / inclination & 35,786~km / $55^{\circ}$ \\
			GEO longitudes & $80^{\circ}$E, $110.5^{\circ}$E, $140^{\circ}$E \\
			FSA / superframe duration & 300~s / 60~s \\
			Slot duration & 3~s \\
			\bottomrule
		\end{tabular}
	\end{table}
	\subsubsection{Scenario and Time Parameters}
	
	The evaluation uses a 24 h BeiDou scenario consisting of
	30 satellites: 24 medium Earth orbit (MEO) satellites, 3 inclined
	geosynchronous orbit (IGSO) satellites, and 3 geostationary orbit
	(GEO) satellites.
	The time organization follows the configuration described in
	Section~\ref{subsec:time_slotted_network}. Each FSA state lasts
	300 s and contains five 60 s superframes, while each superframe is
	divided into twenty 3~s slots.
	The common scenario settings
	are summarized in Table~\ref{tab:experimental_parameters}.
	To mitigate end-of-horizon effects in routing evaluation, the final FSA of the 24-h period is reserved as a guard interval and excluded from the reported routing statistics.
	\subsubsection{Baseline Schemes}
	\label{subsubsec:compared_schemes}
	\paragraph{Topology baseline}
	Fair contact planning (FCP)~\cite{3} serves as the reference ISL topology.
	Our previous work~\cite{3,11} has demonstrated its suitability for onboard GNSS
	operation in terms of computational demand and ranging performance, 
	providing a practical basis for evaluating the additional benefits
	of routing-feedback MTRCC topology construction.
	
	\paragraph{Routing baselines}
	Contact Graph Routing (CGR)~\cite{12,13} and Shortest-Path Tree (SPT) routing~\cite{22,23} are selected as two representative graph-based path-search methods. 
	CGR organizes scheduled communication opportunities in a
	contact graph and searches for time-feasible routes through successive
	contacts. SPT instead operates on a time-varying contact multigraph
	and constructs a shortest-path tree rooted at the source satellite.
	Despite their different graph representations and search organizations,
	both obtain end-to-end routes by exploring future communication
	opportunities. They therefore provide complementary baselines for
	assessing the computational efficiency of DS-SMFP's direct
	forwarding-decision generation.
	Our CGR implementation builds on
	\texttt{pyCGR}\footnote{\url{https://bitbucket.org/juanfraire/pycgr.git}}
	and extends its original single source--single destination search to obtain routes
	to all reachable destinations in one earliest-arrival search from each source.
	SPT is implemented using
	\texttt{pySPSN}\footnote{\url{https://bitbucket.org/olivier-dj/pyspsn}},
	which produces the corresponding source-rooted shortest-path tree.
	In the distributed onboard routing considered here, each satellite
	computes routes from itself to all other satellites. For both CGR and
	SPT, this requires one source-rooted, all-destination search at each
	departure slot over a specified future contact window. The procedure
	is evaluated separately for all satellites, covering the same departure
	slots and source--destination pairs as DS-SMFP on the reference FCP
	topology.
	
	\paragraph{Topology--routing co-optimization baseline}
	To assess the benefit of coupling topology planning with routing performance, a GNSS-adapted Routing-aware FCP (RA-FCP)~\cite{14} is adopted.
	RA-FCP starts from a complete FCP schedule and uses simulated annealing
	to iteratively revise link assignments, evaluating each candidate
	schedule through end-to-end routing performance and contact fairness.
	It therefore introduces routing feedback at the \emph{plan level}.
	In comparison, the proposed MTRCC incorporates future-delivery information
	into link selection at the \emph{slot level}, as each matching is
	constructed. FCP, RA-FCP, and MTRCC are evaluated using the same
	DS-SMFP routing method, enabling their topology-induced communication
	performance to be compared consistently.
	\subsubsection{Computing Platform}
	\label{subsubsec:computing_platform}
	All experiments are conducted on a workstation equipped with an
	AMD Ryzen 7 7700X 8-Core Processor and 32~GB of RAM using Python~3.12.
	
	\begin{table*}[t]
		\centering
		\caption{Computational efficiency under 12 and 24 h planning horizons. Each pair denotes (12 h, 24 h) results.}
		\label{tab:computational_efficiency}
		\footnotesize
		\setlength{\tabcolsep}{7pt}
		\renewcommand{\arraystretch}{1.14}
		\begin{tabular}{@{}lccc@{}}
			\toprule
			\textbf{Method / configuration} &
			\textbf{Topology time (s)} &
			\textbf{Routing time (s)} &
			\textbf{Peak routing working memory (MiB)} \\
			\midrule
			CGR, 30 s window
			& (3.37, 6.67) & (19.71, 36.68) & (0.3117, 0.3117) \\
			CGR, 60 s window
			& (3.37, 6.67) & (60.86, 123.91) & (0.6203, 0.6203) \\
			SPT, 30 s window
			& (3.37, 6.67) & (20.91, 42.44) & (0.3085, 0.3085) \\
			SPT, 60 s window
			& (3.37, 6.67) & (43.04, 88.04) & (0.5615, 0.5615) \\
			\midrule
			DS-SMFP A0: satellite-wise recursion
			& (3.37, 6.67) & (0.14, 0.27) & (0.0227, 0.0227) \\
			DS-SMFP A1: matching-wise joint evaluation + in place
			& (3.37, 6.67) & (0.09, 0.20) & (0.0135, 0.0135) \\
			DS-SMFP A2: A1 + integer delivery slot labels (Final version)
			& (3.37, 6.67) & (0.09, 0.20) & (0.0084, 0.0084) \\
			\midrule
			MTRCC, topology--routing co-construction (MEO1 as master)
			& \multicolumn{2}{c}{\textbf{Integrated:} (13.67, 26.80)} & --- \\
			\bottomrule
		\end{tabular}
		\vspace{2pt}
		\begin{minipage}{0.98\textwidth}
			\scriptsize\emph{Notes:} Peak routing working memory is measured with
			\texttt{tracemalloc}, excluding preloaded topology data and
			accumulated forwarding decisions.
			MTRCC performs integrated topology--routing construction, so
			routing-specific time and working memory are not reported separately.
		\end{minipage}
	\end{table*}
	
	\subsection{DS-SMFP Computational Efficiency}
	\label{subsec:computational_efficiency}
	Table~\ref{tab:computational_efficiency} compares computation time and
	routing working memory over 12 and 24 h planning horizons. CGR, SPT,
	and the DS-SMFP variants use topology schedules generated by
	reverse-order FCP, whereas MTRCC jointly constructs its topology and
	routing policy with MEO1 as the master satellite. 
	For CGR and SPT, distributed onboard execution is emulated by
	performing a single-source, all-destination search for each satellite
	at every departure slot. The reported routing time is the average
	across all 30 satellites, where each satellite's time is accumulated
	over the entire planning horizon. In contrast, one DS-SMFP backward
	sweep computes forwarding decisions for all source--destination pairs.
	Thus, the CGR/SPT results represent the average per-satellite routing
	workload, whereas DS-SMFP reports the computation time required to
	compile the constellation-wide forwarding policy.
	
	Each CGR/SPT search examines a 30 or 60 s future contact window. The
	maximum all-to-all delay under the evaluated FCP topology is
	27.22~s, as reported in Table~\ref{tab:fcp_order_comparison}; 
	hence, a 30 s window is sufficient to
	find an earliest-arrival path for every evaluated source--destination
	state on this topology. A 60 s window provides a more conservative
	setting for contact plans with longer delivery delays. With a fixed
	window, CGR and SPT exhibit approximately linear runtime growth with
	the planning horizon, as does DS-SMFP. However, doubling the search
	window more than triples CGR's 24 h routing time, compared with an
	approximately twofold increase for SPT, indicating greater sensitivity
	of CGR to the enlarged contact-search space in this implementation.
	DS-SMFP instead evaluates STORE and FORWARD decisions using the current
	scheduled matching and the already computed next-slot delivery states.
	Each slot requires $\mathcal{O}(N^2)$ work, without expanding a
	future-contact search graph.
	
	The DS-SMFP ablation starts from the basic satellite-wise backward
	recursion (A0). A1 combines matching-wise joint evaluation, which
	reuses a future-state comparison for both endpoints of a matched ISL,
	with in-place state propagation that removes one time-layer buffer.
	A2 further introduces integer final-delivery-slot labels. These
	successive refinements reduce routing computation time and peak working
	memory, as shown in the table. The latter is measured by
	\texttt{tracemalloc} over active routing computation, excluding
	preloaded topology data and accumulated forwarding decisions.
	Although DS-SMFP computes all-source decisions, a satellite can retain only its own forwarding entries, giving the same per-satellite policy size as the baselines.
	Because MTRCC
	integrates topology selection and routing-state propagation, only its
	joint computation time is reported.
	
	Across all CGR/SPT configurations, the final
	DS-SMFP A2 implementation is over 180 times faster and achieves
	more than a 36-fold reduction in peak routing working memory. 
	Although the Ryzen-based Python test platform differs substantially
	from onboard flight processors, DS-SMFP's shorter measured routing
	times and smaller working memory footprints relative to CGR and SPT
	still substantiate its lightweight computational design. In centralized onboard planning,
	CGR and SPT would additionally repeat source-rooted searches for the
	other satellites, while DS-SMFP would still complete one backward
	sweep, potentially further increasing its relative advantage.
	
	\subsection{Topology and Routing Performance}
	\subsubsection{Forward and Reverse Order FCP Comparison}
	\label{subsubsec:fcp_order_comparison}
	\begin{table}[t]
		\centering
		\caption{Ranging and all-to-all routing performance under
			forward- and reverse-order FCP.}
		\label{tab:fcp_order_comparison}
		\footnotesize
		\setlength{\tabcolsep}{4pt}
		\renewcommand{\arraystretch}{1.12}
		\begin{tabular}{@{}p{0.53\columnwidth}rr@{}}
			\toprule
			\textbf{Metric} & \textbf{Forward} & \textbf{Reverse} \\
			\midrule
			Mean ranging partners           & 15.9370 & 15.9370 \\
			Minimum ranging partners        & 10      & 10      \\
			Mean PDOP                       & 1.0006  & 1.0006  \\
			\addlinespace[2pt]
			Mean all-to-all delay (s)       & 10.8826 & 10.8824 \\
			Maximum all-to-all delay (s)    & 27.2234 & 27.2276 \\
			Mean all-to-all hop count       & 2.5721  & 2.5717  \\
			Maximum all-to-all hop count    & 8       & 8       \\
			\bottomrule
		\end{tabular}
		\vspace{2pt}
		\parbox{\columnwidth}{\scriptsize\emph{Note:} Ranging partners are
			distinct neighbors within a 60 s superframe.}
	\end{table}
	MTRCC constructs matchings in reverse time order because its link
	selection uses already computed future-delivery states. Accordingly,
	reverse-order FCP serves as the ranging-oriented baseline topology in
	subsequent comparisons. Conventional FCP selects matchings
	chronologically while updating history-dependent fairness weights.
	Reversing the phase order retains the same fairness-driven matching
	selections within each FSA but changes their placement on the physical
	time axis. We assess this effect by generating both schedules under
	identical visibility and matching constraints and evaluating their
	routing performance with DS-SMFP.
	
	Table~\ref{tab:fcp_order_comparison} shows that both schedules provide
	identical ranging performance: an average of 15.937 distinct ranging
	partners per satellite per 60 s superframe, a minimum of 10, and a
	mean PDOP of 1.0006. 
	For ISL ranging, PDOP characterizes the geometry of the distinct
	satellites contacted by a given satellite within each superframe.
	The reported mean PDOP is averaged over all satellites and superframes;
	lower values indicate more favorable ranging geometry.
	
	Their all-to-all routing performance is likewise
	nearly identical, with differences of only 0.20~ms in mean delay and
	4.20~ms in maximum delay, negligible changes in mean hop count, and
	the same maximum of eight hops. Thus, reverse-order FCP preserves the
	ranging performance of chronological FCP while introducing negligible
	communication differences in the evaluated scenario. It is therefore
	used as the baseline topology for the following experiments, so that
	the gains of MTRCC can be attributed to topology--routing
	co-construction rather than to reversing the FCP schedule alone.
	\subsubsection{DS-SMFP Routing Correctness Verification}
	\label{subsubsec:routing_correctness}
	\begin{figure}[t]
		\centering
		\begin{minipage}[t]{\linewidth}
			\centering
			\includegraphics[width=0.7\linewidth]{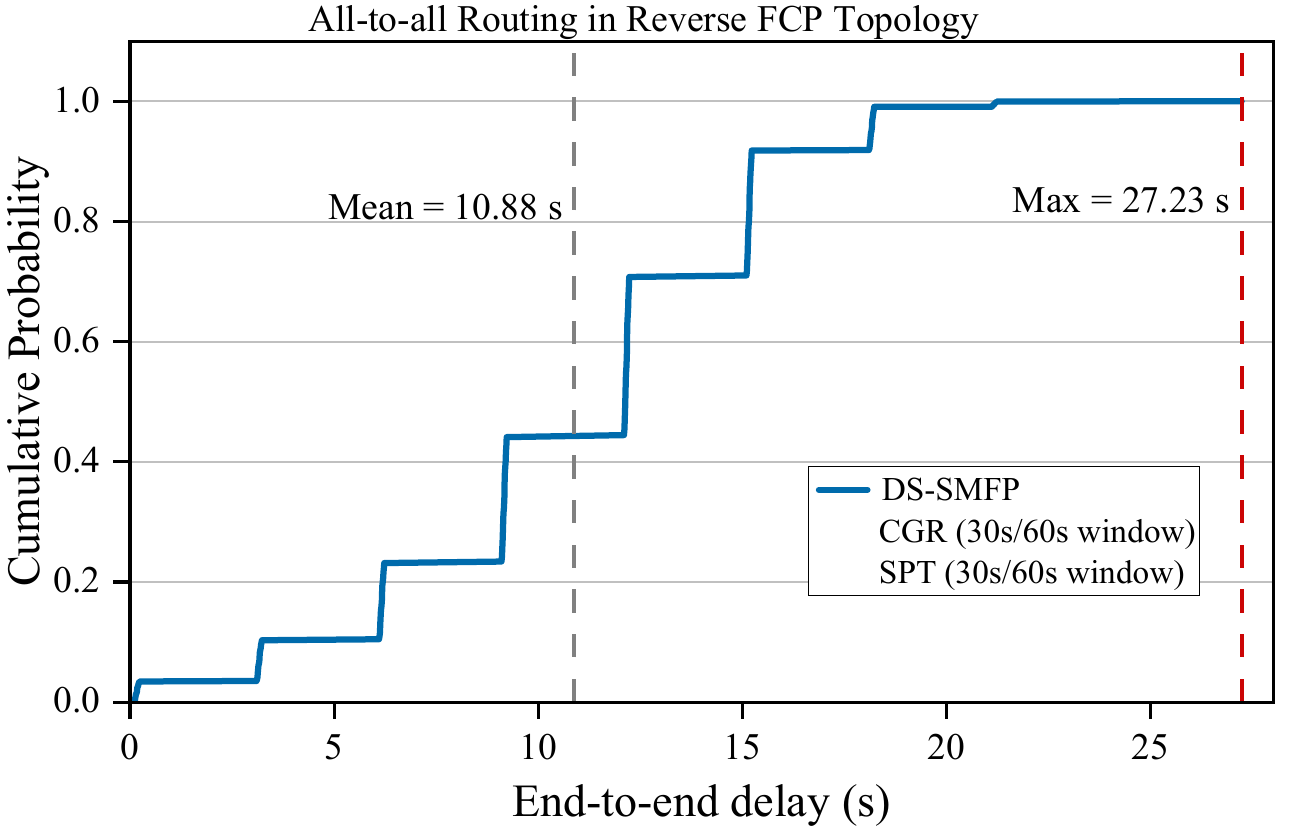}
			\vspace{-1mm}
			
			{\footnotesize (a) End-to-end delay}
		\end{minipage}
		
		\vspace{1mm}
		\begin{minipage}[t]{\linewidth}
			\centering
			\includegraphics[width=0.7\linewidth]{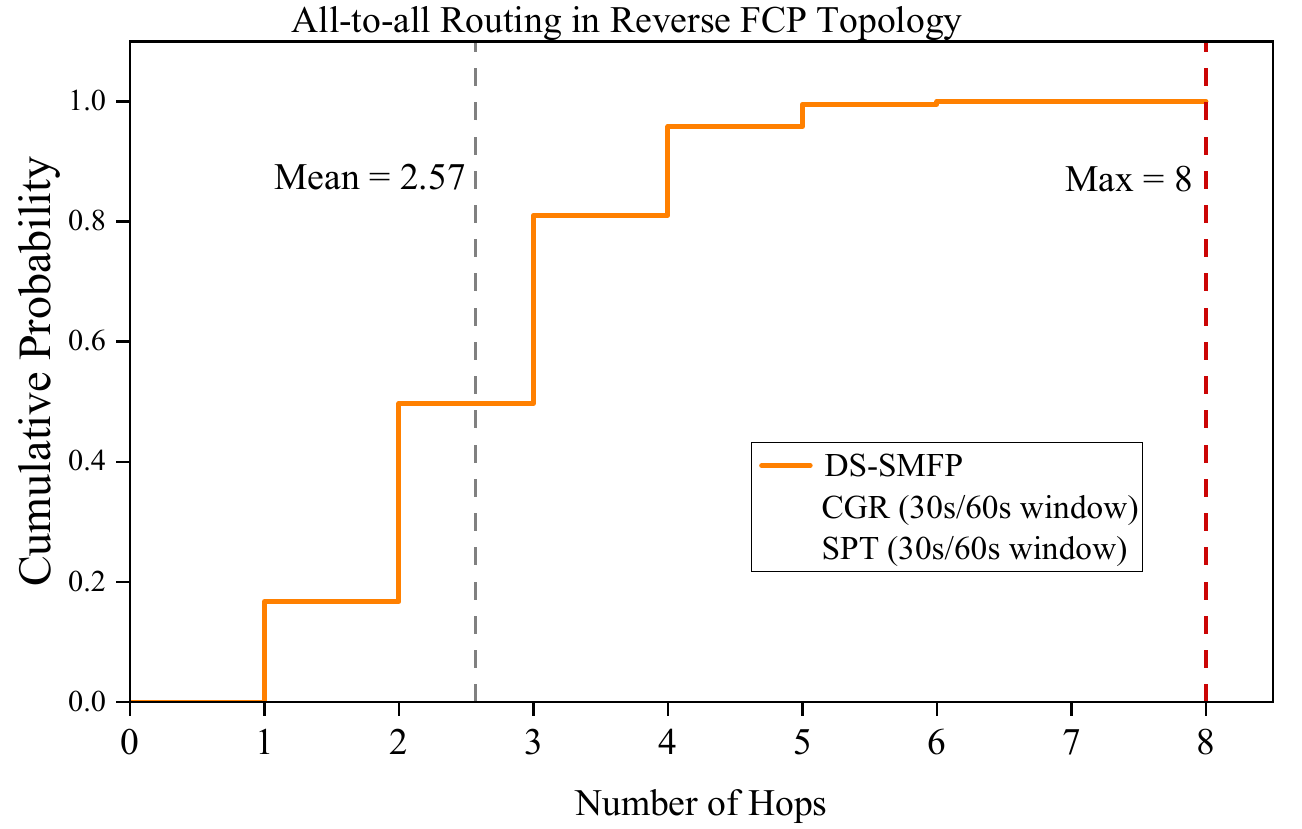}
			\vspace{-1mm}
			
			{\footnotesize (b) Hop count}
		\end{minipage}
		
		\caption{All-to-all routing on the reverse order FCP topology.
			CGR and SPT with 30 and 60 s look-ahead windows produce the same
			empirical delay and hop-count CDFs as DS-SMFP. Each panel therefore
			shows one common curve for all five configurations.}
		\label{fig:routing_correctness_cdf}
	\end{figure}
	The routing correctness of DS-SMFP is verified against CGR and SPT on an identical reverse-order FCP topology.
	All three methods seek the
	earliest arrival, with minimum hop count as the secondary criterion,
	under the capacity-sufficient store-and-forward model. Their routing
	results are therefore expected to agree when the search windows contain
	the earliest-arrival paths. Fig.~\ref{fig:routing_correctness_cdf}
	compares the all-to-all end-to-end delay and hop-count distributions
	of DS-SMFP, CGR, and SPT, with 30 and 60 s windows for the latter
	two methods. All five configurations yield identical empirical CDFs.
	This agreement provides an empirical correctness check for DS-SMFP's
	backward recurrence and confirms that both search-window settings
	achieve the same observed routing performance on the evaluated FCP
	schedule. The computational gains reported above thus arise from
	DS-SMFP's routing organization rather than different delivery outcomes
	on a fixed topology.
	Accordingly, DS-SMFP is adopted as the common
	routing method for evaluating end-to-end communication performance
	across different topology schemes in the subsequent experiments.

	\subsubsection{MTRCC Ranging--Communication Trade-off}
	\label{subsubsec:ranging_communication_tradeoff}
	\begin{figure*}[t]
		\centering
		\includegraphics[width=0.90\textwidth]{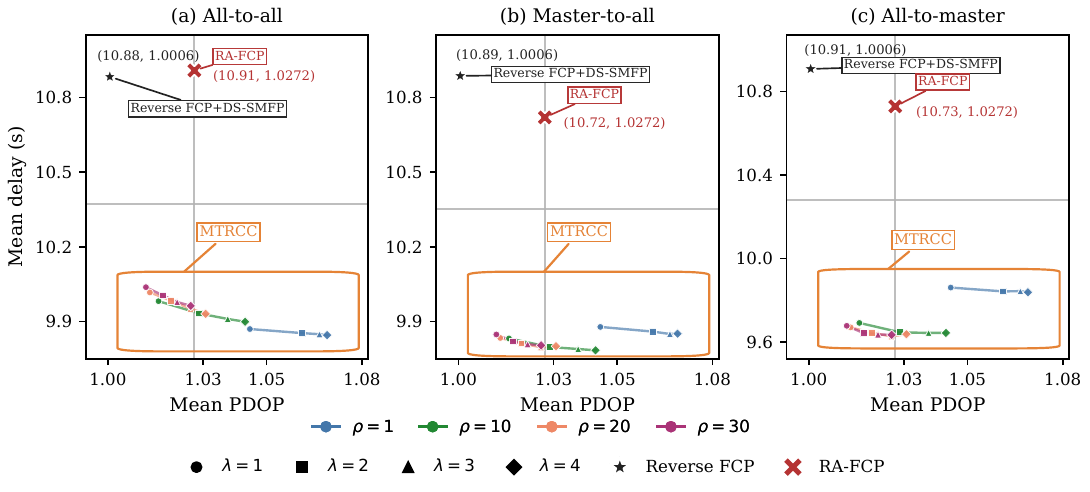}
		\caption{Ranging--communication trade-off with MEO1 as the master
			satellite: mean PDOP versus mean delay for (a) all-to-all,
			(b) master-to-all, and (c) all-to-master traffic.}
		\label{fig:mtrcc_ranging_tradeoff}
	\end{figure*}
	\label{subsubsec:end_to_end_distributions}
	\begin{figure*}[t]
		\centering
		\includegraphics[width=0.9\textwidth]{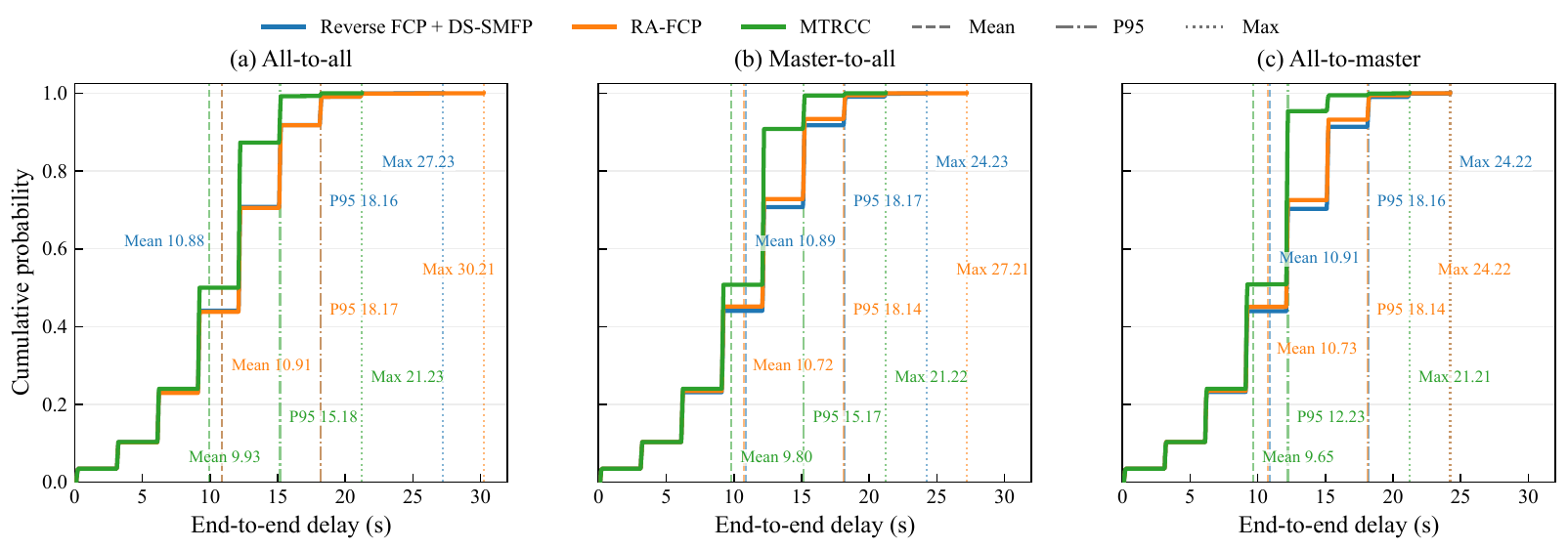}
		\caption{End-to-end delay distributions under reverse-order FCP,
			RA-FCP, and MTRCC ($\lambda=2$, $\rho=10$): (a) all-to-all,
			(b) master-to-all, and (c) all-to-master traffic. Vertical reference
			lines indicate the mean, 95th percentile (P95), and maximum delay
			for each topology.}
		\label{fig:mtrcc_delay_cdf}
	\end{figure*}
	
	\begin{figure*}[t]
		\centering
		\includegraphics[width=0.9\textwidth]{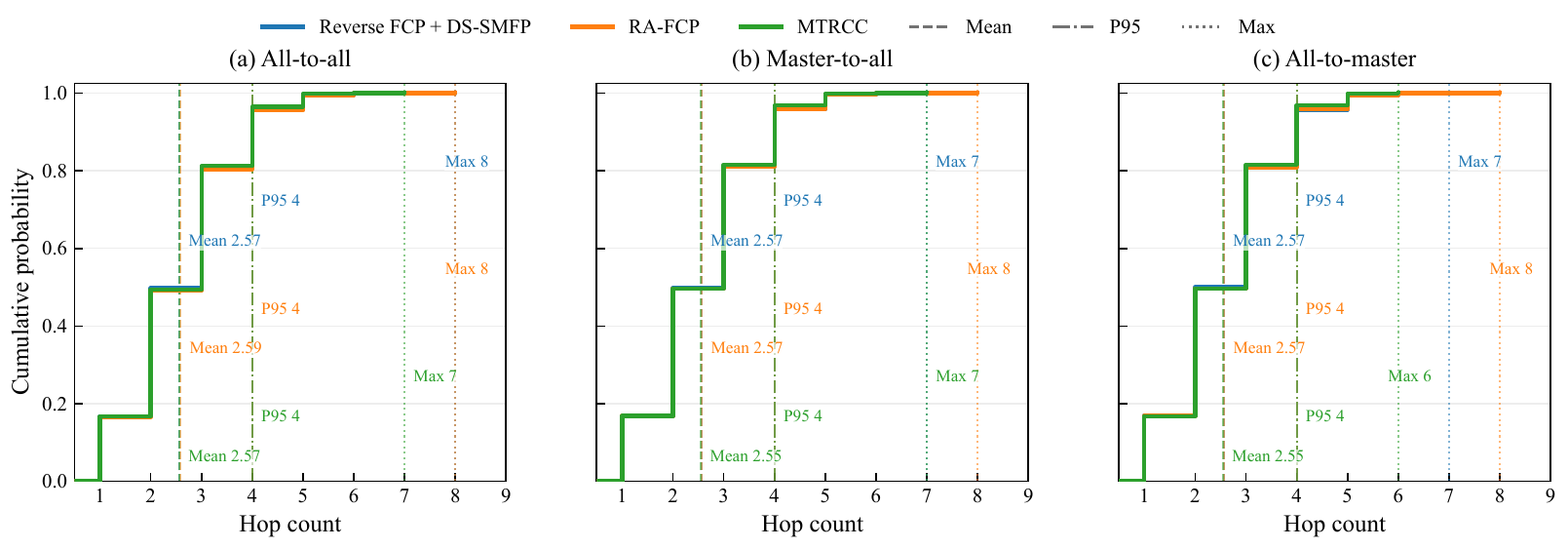}
		\caption{End-to-end hop-count distributions for the same three
			topologies and traffic directions as in Fig.~\ref{fig:mtrcc_delay_cdf}.
		}
		\label{fig:mtrcc_hop_cdf}
	\end{figure*}
	The routing-aware matching objective in \eqref{eq:joint_edge_weight}
	allows communication performance to be improved while retaining the
	ranging-oriented fairness of FCP. We examine this trade-off with MEO1
	as the master satellite, comparing reverse-order FCP topology with DS-SMFP routing,
	GNSS-adapted RA-FCP, and MTRCC. For MTRCC, the
	routing-feedback coefficient is varied over
	$\lambda\in\{1,2,3,4\}$ and the master-priority factor over
	$\rho\in\{1,10,20,30\}$. Fig.~\ref{fig:mtrcc_ranging_tradeoff}
	plots mean positioning dilution of precision (PDOP) against mean
	end-to-end delay for all-to-all (AA), master-to-all (MA), and
	all-to-master (AM) traffic. All schemes use DS-SMFP for the final
	routing evaluation.
	
	For a comparison with plan-level routing feedback, the RA-FCP baseline optimizes a weighted combination of master-prioritized mean delivery delay, overall network mean delay, and the Jain fairness index of scheduled contacts, with coefficients of 0.8, 0.1, and 0.1, respectively.
	Its simulated-annealing
	search generates neighboring plans by randomly selecting FSA states
	and scheduling phases, removing subsets of their current ISLs, and
	recomputing the corresponding matchings. Each candidate is then
	assessed against the routing and contact-fairness objective. The
	search is run for 300,000 iterations and takes 28740~s in this
	experiment.
	As shown in Fig.~\ref{fig:mtrcc_ranging_tradeoff}, reverse-order FCP
	achieves the lowest mean PDOP of $1.0006$, consistent with its
	exclusive emphasis on contact fairness. Its AA, MA, and AM mean
	delays are $10.88$, $10.89$, and $10.91$~s, respectively. RA-FCP
	increases mean PDOP to $1.0272$ and reduces the two master-related
	delays to $10.72$ and $10.73$~s, but its AA delay rises slightly to
	$10.91$~s. 
	The plan-level optimization therefore shifts communication
	performance toward the prioritized master-related traffic, with a
	limited gain in the two directions and a small cost to overall network
	delay. By contrast, the MTRCC configurations occupy a substantially
	lower-delay region in all three panels, improving AA as well as both
	master-related traffic directions.
	
	The two MTRCC parameters control different aspects of this trade-off.
	For a fixed $\rho$, increasing $\lambda$ generally decreases routing
	delay while increasing PDOP, as more weight is assigned to the future
	delivery benefit of candidate links. For a fixed $\lambda$, increasing
	$\rho$ shifts the communication gain toward the master: MA and AM
	delays decrease, whereas AA delay tends to increase. The largest
	changes are observed when $\rho$ rises from $1$ to $10$, with smaller
	additional changes beyond $10$. Thus, $\lambda$ primarily balances
	ranging geometry against communication delay, while $\rho$ adjusts the
	allocation of communication benefit between network-wide and
	master-related traffic.
	
	All tested topologies retain at least nine distinct ranging partners
	per satellite in every superframe, providing diverse measurements for
	autonomous orbit determination. Although stronger routing weights
	increase PDOP, the mean values remain well below $1.5$ across the
	parameter scan, indicating favorable ranging geometry. Thus, MTRCC
	maintains both measurement diversity and geometric quality while
	reducing communication delays. In the master-satellite architecture,
	lower AM delay accelerates the aggregation of ranging for onboard
	orbit determination, whereas lower MA delay facilitates the subsequent
	dissemination of updated orbit and navigation information. These
	complementary benefits make the resulting link schedules well suited
	to the ranging and information-exchange requirements of autonomous
	GNSS operation.
	
	The difference from RA-FCP is also structural.
	Its randomized plan-level neighborhood generation does not use the
	prospective routing gain of individual candidate links to direct the
	rematching process, making the search costly over a rapidly changing
	GNSS contact schedule. MTRCC instead evaluates that gain as each
	slot-level matching is constructed, requiring $26.80$~s for the
	integrated 24-h topology--routing computation in
	Table~\ref{tab:computational_efficiency}. These results demonstrate
	that construction-stage routing feedback offers an effective and
	tunable ranging--communication trade-off at substantially lower
	computational cost in the evaluated implementation.
	
	\subsubsection{MTRCC End-to-End Performance Distributions}
	
	To examine the communication gains of MTRCC beyond mean-delay metrics,
	we select a representative configuration with $\lambda=2$ and
	$\rho=10$, using MEO1 as the master satellite. Its end-to-end delay
	and hop-count distributions are compared with those of reverse-order
	FCP and RA-FCP for AA, MA, and AM traffic. This comparison examines
	whether the delay improvements extend to the distribution tails and
	how they relate to changes in path length.
	
	As shown in Fig.~\ref{fig:mtrcc_delay_cdf}, MTRCC shifts the delay
	CDFs toward earlier delivery in all three traffic directions. Its
	AA, MA, and AM mean delays are $9.93$, $9.80$, and $9.65$~s,
	respectively, compared with $10.88$, $10.89$, and $10.91$~s for
	reverse-order FCP. The improvement extends to the distribution tails:
	MTRCC reduces the corresponding P95 delays to $15.18$, $15.17$,
	and $12.23$~s, from approximately $18.2$~s under both baselines.
	The maximum delays are also reduced to $21.23$, $21.22$, and
	$21.21$~s, respectively. RA-FCP improves the two master-related mean
	delays but slightly increases the AA mean to $10.91$~s and the AA/MA
	maximum to $30.21$/$27.21$~s, indicating less consistent benefits across
	traffic directions and extreme routing states.
	
	Fig.~\ref{fig:mtrcc_hop_cdf} shows more modest changes in path length.
	All three methods retain a P95 of four hops in every traffic direction,
	while MTRCC reduces the maximum AA hop count from eight to seven and
	the maximum AM hop count from seven under reverse-order FCP and eight
	under RA-FCP to six. The maximum MA hop count remains seven.
	The relatively small hop-count changes, alongside the pronounced delay
	reductions, suggest that MTRCC gains much of its advantage by scheduling
	useful forwarding opportunities earlier and reducing inter-slot waiting,
	rather than primarily by shortening end-to-end paths. This benefits
	both measurement aggregation toward the master and subsequent
	information dissemination from it.
	\subsubsection{MTRCC Master-Satellite Generality}
	\label{subsubsec:master_satellite_generality}
	\begin{table*}[t]
		\centering
		\caption{Master-satellite generality across MEO, IGSO, and GEO. Delay and hop-count entries report mean/maximum values.}
		\label{tab:master_satellite_generality}
		\footnotesize
		\setlength{\tabcolsep}{4pt}
		\renewcommand{\arraystretch}{1.12}
		\begin{tabular}{@{}llccrrrc@{}}
			\toprule
			& & & & \multicolumn{3}{c}{\textbf{End-to-end delay (s)}} & \textbf{AA hops} \\
			\cmidrule(lr){5-7}
			\textbf{Master} & \textbf{Method} & \textbf{Mean PDOP} & \textbf{Min. links} &
			\textbf{AA mean/max} & \textbf{MA mean/max} & \textbf{AM mean/max} & \textbf{Mean/max} \\
			\midrule
			\multirow{3}{*}{MEO1}
			& Reverse FCP & 1.0006 & 10 & 10.88/27.23 & 10.89/24.23 & 10.91/24.22 & 2.57/8 \\
			& RA-FCP      & 1.0272 &  9 & 10.91/30.21 & 10.72/27.21 & 10.73/24.22 & 2.59/8 \\
			& MTRCC       & 1.0308 & 10 &  9.931/21.23 &  9.801/21.22 &  9.638/21.15 & 2.56/7 \\
			\addlinespace[3pt]
			\multirow{3}{*}{IGSO1}
			& Reverse FCP & 1.0006 & 10 & 10.88/27.23 & 10.92/27.23 & 10.96/27.17 & 2.57/8 \\
			& RA-FCP      & 1.0288 & 10 & 10.91/27.23 & 10.74/24.22 & 10.77/24.23 & 2.59/8 \\
			& MTRCC       & 1.0292 &  9 &  9.929/21.23 &  9.853/21.23 &  9.685/21.23 & 2.57/7 \\
			\addlinespace[3pt]
			\multirow{3}{*}{GEO1}
			& Reverse FCP & 1.0006 & 10 & 10.88/27.23 & 10.88/27.20 & 10.85/27.22 & 2.57/8 \\
			& RA-FCP      & 1.0240 &  9 & 10.90/27.23 & 10.74/24.23 & 10.71/24.23 & 2.59/8 \\
			& MTRCC       & 1.0289 & 10 &  9.930/21.23 &  9.838/21.23 &  9.681/21.23 & 2.57/7 \\
			\bottomrule
		\end{tabular}
		\vspace{2pt}
		\parbox{0.98\textwidth}{\scriptsize\emph{Note:} Min. links denotes the minimum number of distinct ranging partners per satellite per superframe. Reverse FCP uses the same topology in all three groups; its MA and AM metrics depend on the selected master. All routing delays include physical one-way light time.}
	\end{table*}
	
	To assess MTRCC across orbital classes, we select MEO1, IGSO1, and GEO1
	as representative MEO, IGSO, and GEO master satellites. MTRCC uses
	$\lambda=4$ and $\rho=20$ in all three cases, complementing the
	$(2,10)$ setting used in the preceding analysis. Both $(4,20)$ and $(2,10)$ settings
	represent balanced points in the ranging--communication trade-off
	shown in Fig.~\ref{fig:mtrcc_ranging_tradeoff}, providing substantial
	communication-delay reductions while retaining favorable PDOP. RA-FCP is
	optimized separately for each master using the same weighted
	simulated-annealing objective as in the preceding trade-off
	experiment, whereas reverse-order FCP uses one master-independent
	topology; its MA and AM delays are evaluated for each master.

	Table~\ref{tab:master_satellite_generality} shows that MTRCC
	consistently improves communication for all three master choices.
	It reduces the all-to-all (AA) mean delay from 10.88~s under
	reverse-order FCP to 9.93~s, while master-to-all (MA) and
	all-to-master (AM) mean delays reach 9.80--9.85~s and
	9.64--9.68~s, respectively. These master-related delays are
	also lower than those of RA-FCP. The maximum AA delay decreases
	from 27.23 to 21.23~s, and the maximum AA hop count from eight
	to seven. Meanwhile, all three MTRCC schedules retain at least
	nine distinct ranging partners per satellite per superframe,
	with mean PDOP between 1.0289 and 1.0308.
	
	The small differences among the three master choices are consistent
	with their distinct visibility conditions. Across the 24 h
	five minute FSA states, MEO1 has an average of 19.54 visible
	satellites, compared with 12.21 for IGSO1 and 12.10 for GEO1.
	MEO1 consequently offers more candidate ISLs for matching selection,
	which is consistent with its slightly lower MA and AM delays.
	The improvements observed for IGSO1 and GEO1, despite their smaller
	visible-neighbor sets, indicate that MTRCC can support
	bidirectional master-oriented communication across the various
	orbital classes while preserving ranging diversity and geometry.

	\subsection{Further Discussion}
	The following discussion examines the adaptability of the proposed methods to practical ISL operation and broader topology--routing scenarios.
	\label{subsec:evaluation_discussion}
	\subsubsection{DS-SMFP Compatibility with Intra-Slot Tx/Rx Scheduling}
	To mitigate interference between simultaneously operated transmit and
	receive beams, BeiDou ISLs divide each 3 s time slot into two 1.5 s
	transmit (Tx) and receive (Rx) intervals. The endpoints of a scheduled
	ISL adopt complementary Tx/Rx states, allowing communication in the
	corresponding direction. DS-SMFP determines the binary STORE/FORWARD
	flag $B_k(i,d)$ at the slot level. When compiling the onboard forwarding
	table, this flag can be combined by a Boolean AND with the local Tx
	state (1 for Tx and 0 for Rx) in each interval, so forwarding is
	executed only when the satellite is scheduled to transmit.
	This intra-slot division preserves the integer delivery-slot ordering
	underlying DS-SMFP. Since each satellite has only one scheduled peer
	per slot, a packet received by an intermediate relay cannot be
	forwarded to a different satellite until a later slot. Provided that
	scheduled transfers finish within their assigned intervals, the final
	delivery slot $R_k(i,d)$ remains unchanged. The Tx/Rx order only adds
	an offset of at most 1.5~s to the slot-start arrival-time approximation
	in \eqref{eq:arrival_from_slot_label}. For a given destination and
	final-delivery slot, the scheduled matching fixes the incoming peer
	and its receive interval, so all competing paths share the same
	within-slot offset. Consequently, the intra-slot Tx/Rx division does
	not alter the earliest-arrival or minimum-hop routing decisions at the
	slot level, and the forwarding paths computed by DS-SMFP remain unchanged.
	
	\subsubsection{Topology-Routing Co-Construction Applicability Beyond MTRCC}
	Topology--routing co-construction uses future-delivery states to assess
	the routing value of candidate ISLs and is not restricted to FCP-based
	matching or master-oriented traffic. In the evaluated MTRCC
	implementation, this routing feedback is combined with FCP's
	ranging-fairness scores and source--destination weights that prioritize
	master-related communication. Other topology constructors can incorporate
	the same routing feedback into their link-selection objectives, provided
	that they generate feasible topology in reverse time order. Likewise,
	the source--destination weights can reflect other communication
	requirements, such as network-wide information dissemination, data
	aggregation at multiple satellites, or prioritized exchanges between
	selected satellite groups. These adaptations change the matching
	objective and traffic priorities while retaining the DS-SMFP backward
	state propagation itself.

	\section{Conclusion}
	\label{sec:conclusion}
	
	Autonomous GNSS networking requires efficient onboard routing over rapidly changing, single-peer per slot ISLs. This paper developed DS-SMFP, a lightweight routing method that exploits the scheduled matching structure and predictable propagation delays arising from GNSS orbital geometry.
	Through destination-seeded backward propagation, DS-SMFP directly generates earliest-arrival forwarding decisions.
	In a 24-h BeiDou evaluation, it reproduces the end-to-end delay and
	hop-count distributions of CGR and SPT while running over 180 times faster and using over 36 times less peak routing working memory across the tested configurations.
	
	The one-slot dependency of DS-SMFP also makes future-delivery information available during topology construction. MTRCC uses this
	information to incorporate routing gains into FCP's ranging-oriented
	matching objective, completing the integrated 24-h topology and routing computation in 26.80~s. Compared with reverse FCP, a representative
	MTRCC configuration reduces mean all-to-all delay from 10.88 to 9.93~s, master-to-all delay from 10.89 to 9.80~s, and all-to-master delay from
	10.91 to 9.64~s, with only a modest reduction in ranging performance.
	Its tunable ranging--communication trade-off and consistent communication improvements across MEO, IGSO, and GEO master satellites further support its applicability to autonomous GNSS operation.
	
	Autonomous GNSS operation depends on both diverse inter-satellite ranging measurements and timely delivery of measurement data and updated navigation information throughout the constellation. DS-SMFP and MTRCC address these connected requirements through onboard forwarding computation and
	routing-informed ISL scheduling. Together, they extend GNSS autonomy
	beyond link scheduling toward coordinated topology and routing decisions, strengthening the self-sufficiency of the navigation constellation.

	\bibliographystyle{IEEEtran}
	\bibliography{references}
	
\end{document}